\documentclass[format=acmsmall, review=false, screen=true]{acmart}

\usepackage{booktabs} % For formal tables
\usepackage{graphicx}
\usepackage{array}
\usepackage{url}
\usepackage{fancybox}
\usepackage{multirow}
\usepackage{color}
\usepackage{colortbl}
\usepackage{fancyhdr}
\usepackage{subfig}
\usepackage{diagbox}
\usepackage{bbding}
\usepackage{placeins}
\usepackage{balance}
\usepackage{listings}

\usepackage[linesnumbered, ruled,vlined]{algorithm2e}

\setcopyright{acmcopyright}
\acmJournal{TOSEM}
\acmYear{2021} \acmVolume{1} \acmNumber{1} \acmArticle{1} \acmMonth{1} \acmPrice{15.00}\acmDOI{10.1145/3488245}
\newcommand{\todo}[1]{}
\renewcommand{\todo}[1]{{\color{red} TODO: {#1}}}

\begin{document}
	\title{Why Do Smart Contracts Self-Destruct? \\ Investigating the Selfdestruct Function on Ethereum}
	\titlenote{Corresponding Author: Xin Xia}
	\author{Jiachi Chen}
	%\authornote{XXXX}
	\affiliation{%
		\institution{Monash University}
	}
	\email{jiachi.chen@monash.edu}

	\author{Xin Xia}
	%\authornote{XXXX}
	\affiliation{%
		\institution{Monash University}
	}
	\email{xin.xia@acm.org}
	
	\author{David Lo}
	%\authornote{XXXX}
	\affiliation{%
		\institution{Singapore Management University}
	}
	\email{davidlo@smu.edu.sg}
	
	\author{John Grundy}
	%\authornote{XXXX}
	\affiliation{%
		\institution{Monash University}
	}
	\email{John.Grundy@monash.edu}
	
%	\renewcommand{\shortauthors}{Jiachi et al.}

		\begin{abstract}
	%Smart contracts cannot be modified after deploying them to the Ethereum blockchain. Executing the \textit{selfdestruct} function is the only way to remove smart contracts from Ethereum. 
	The \textit{selfdestruct} function is provided by Ethereum smart contracts to destroy a contract on the blockchain system. However, it is a double-edged sword for developers. On the one hand, using \textit{selfdestruct} function enables developers to remove smart contracts (SC) from Ethereum and transfers Ethers when emergency situations happen, e.g. being attacked. On the other hand, this function can increase the complexity for the development and open an attack vector for attackers. To better understand the reasons why SC developers include or exclude the \textit{selfdestruct} function in their contracts, we conducted an online survey to collect feedback from them and summarize the key reasons. Their feedback shows that 66.67\% of the developers will deploy an updated contract to the Ethereum after destructing the old contract. According to this information, we propose a method to find the self-destructed contracts (also called predecessor contracts) and their updated version (successor contracts) by computing the code similarity. By analyzing the difference between the predecessor contracts and their successor contracts, we found five reasons that led to the death of the contracts; two of them (i.e., \textit{Unmatched ERC20 Token and Limits of Permission})  might affect the life span of contracts. We developed a tool named \textsc{LifeScope} to detect these problems. \textsc{LifeScope} reports 0 false positives or negatives in detecting  \textit{Unmatched ERC20 Token}. In terms of \textit{Limits of Permission}, \textsc{LifeScope} achieves 77.89\% of F-measure and 0.8673 of AUC in average. According to the feedback of developers who exclude \textit{selfdestruct} functions, we propose suggestions to help developers use \textit{selfdestruct} functions in Ethereum smart contracts better. 
	
\end{abstract} 
	
	%	\begin{CCSXML}
	%		<ccs2012>
	%		<concept>
	%		<concept_id>10010520.10010553.10010562</concept_id>
	%		<concept_desc>Software and its engineering</concept_desc>
	%		<concept_significance>500</concept_significance>
	%		</concept>
	%		<concept>
	%	\end{CCSXML}
	
	%	\ccsdesc[500]{Software and its engineering~Software defect analysis}


	\maketitle
	\section{Introduction}
\label{Introduction}
%1. introduce blockchain market
The great success of Bitcoin~\cite{bitcoin} shows the enormous potential of blockchain technology~\cite{blockchain-wiki}. People usually regard Bitcoin as a representative of blockchain 1.0~\cite{xiaoqiSurvey}, the first generation of blockchain technology. In blockchain 1.0, the blockchain technology is usually used to make cryptocurrency~\cite{Cryptocurrency-wiki}, e.g., Bitcoin, Ripple Coin~\cite{ripple}. The usage scenario of cryptocurrencies in blockchain 1.0 is limited, as the main application for them is storing and transferring values~\cite{efanov2018all}. 

%2. introduce smart contract and Ethereum
The birth of Ethereum~\cite{ethereum}  changed this situation at the end of 2015. Ethereum leverages a technology named \textit{smart contracts}, which can be regarded as a program that runs on the blockchain. Smart contracts are usually developed in a high-level programming language, e.g., Solidity~\cite{Solidity}. The blockchain technology provides an  immutability feature for smart contracts, which means all of the smart contracts are self-executed and can't be modified. Even the creator of the contract cannot modify the code after deploying the contract to the Ethereum. By utilizing a smart contract, developers can easily design their DApps (decentralized applications)~\cite{dapp}. The appearance of Ethereum marked the point that blockchain technology upgraded from blockchain 1.0 to blockchain 2.0. By Sept. 2019, millions of smart contracts on the Ethereum~\cite{EtherScan} have been applied to different fields, such as gaming~\cite{cryptokitties} and monetization~\cite{erc20}, with many other application domains under exploration.

%3. However, smart contract is not safe. Introducing DAO attack and selfdestruct function
However, the features of Ethereum also make it easy to be attacked. First, Ethereum is a permission-less network; smart contracts on Ethereum can be executed by everyone, including attackers. Second, all the data stored on the blockchain, transactions, and bytecode of smart contracts are visible to the public, which makes smart contracts become attractive targets for attackers. In 2016, attackers utilized a vulnerability (reentrancy~\cite{oyente}) to attack a smart contract owned by an organization named DAO (Decentralized Autonomous Organization). This attack made the organization lose 3.6 million Ethers\footnote{Ether is the cryptocurrency generated by the Ethereum platform. An Ether worth \$1270 on Jan. 2021.}. People usually call this attack a DAO attack~\cite{DAOAttack}.  Actually, the attack continued for several days and the organization even noticed that their contract had been attacked at that time. However, they could not stop the attack or transfer the Ethers because of the immutability feature of smart contracts. 

This DAO attack attracted great attention from both academia and industry. Some previous works~\cite{bestPractice, smellDefinition} advice contracts add some mechanisms to stop the contracts or transfer the Ethers when emergency situations happen, e.g., a contract being attacked. In this case, the owners can reduce the impact of financial loss. Solidity provides a novel \textit{selfdestruct} function~\cite{Solidity}. By calling this \textit{selfdestruct} function, a smart contract can be removed from the blockchain and all the Ethers on the contract will be transfered to a specified address, which is an unique identification for the account\footnote{There are two types of accounts on Ethereum, i.e., External Owned Account (EOA)  and contract account. EOA is controlled by users. Contract account is controlled by its code.}  on Ethereum. 

This \textit{selfdestruct} function is however a double-edged sword for developers. On the one hand, the function enables contract owners have the ability to reduce financial loss when emergency situations happen. On the other hand, this function is also harmful. The function might open an attack vector for attackers. It may also lead to a trust concern from the contract users, as the contract owners can transfer user's Ethers that are stored on the contract. These conflicting features make the \textit{selfdestruct} function valuable to be investigated.\textbf{ In this paper, we call a smart contract that has executed the \textit{selfdestruct} function as a `self-destructed contract'. } To better understand the developers' perspective about this unique Ethereum smart contract \textit{selfdestruct} function, we designed an online survey to collect their opinions, and to help us to answer the following research question: 

%In our analysis, we found that 2,786 (5.1\%) of smart contracts among 54,739 contracts we crawled from Etherscan~\cite{EtherScan} contain a \textit{selfdestruct} function.  

\noindent{ \textbf{\textit{RQ1: Why do smart contract developers include or exclude \textit{selfdestruct} functions in their contracts?}}} 

We sent our survey to 996 smart contract developers and received 88 responses. Their feedback shows that there are six reasons why developers exclude the \textit{selfdestruct} function. The top two most popular reasons are \textit{security concerns} and \textit{trust concerns}. Developers are  worried that the \textit{selfdestruct} function in their contract will open an attack vector for attackers. Besides, this function can also reduce users' confidence of the contract as the contract owner have ability to transfer users' Ethers that stored on the contract balance. To address these concerns of developers, we provide six suggestions in Section~\ref{lab:discussion}, which can help developers to better use the \textit{selfdestruct} function. In terms of why developers include a \textit{selfdestruct} function, our survey feedback shows that two thirds of developers will kill their contracts when security vulnerabilities are found, or if they want to upgrade their smart contract's functionalities. After fixing bugs or upgrading the contracts, they will deploy a new version of the contract. This finding inspired our second research question:

\noindent{ \textbf{\textit{RQ2: Why do smart contracts on Ethereum self-destruct?}}} 

We called the \textit{self-destructed} contract as a \textit{`Predecessor'} contract, and its upgraded version as a \textit{'Successor'} contract. By comparing the difference between \textit{Predecessor} contract and \textit{Successor} contract, we can identify the reasons why contract destructed, e.g., security reasons. We propose a method that leverages a clone detection tool (\textsc{SmartEmbed}~\cite{smartembed, gao2020checking}) to find \textit{Predecessor} contracts and their  \textit{Successor} contracts. Then, we summarize 5 common reasons why contracts destructed by conducting \textit{open card sorting}~\cite{spencer2009card}. 

As a result, we summarize 5 common self-destruct reasons, detailed in Table~\ref{tab:reasons}. Two of them -- \textit{Unmatched ERC20 Token} and \textit{Limits of Permission} -- might affect the life span of contracts. Therefore, an automatic tool to detect these problems would be helpful to extend the life span of smart contracts. This motivated us to investigate our third research question: 

\noindent{ \textbf{\textit{RQ3: How can we detect lifespan-based smart contract problems automatically?}}}

We designed a tool named \textsc{LifeScope}, which can be used to detect \textit{Unmatched ERC20 Token} and \textit{Limits of Permission} problems. For \textit{Unmatched ERC20 Token},  \textsc{LifeScope} uses ASTs (Abstract Syntax Trees) to parse the source code and extract related information. \textsc{LifeScope} obtains 100\% of F-measure for detecting this problem. For \textit{Limits of Permission}, \textsc{LifeScope} first transfers code to a TF-IDF representation and utilizes a machine learning method to predict the permission.  \textsc{LifeScope} achieves an F-measure and AUC of 77.89\% and 0.8673 for this task.

The main contributions of this paper are: 
\begin{itemize}\setlength{\itemsep}{3pt}
	\item To the best of our knowledge, this is the most comprehensive empirical work that investigates the \textit{selfdestruct} function of smart contracts in Ethereum. We conduct an online survey to collect feedback from developers. According to this survey feedback, we summarize 6 reasons why developers add \textit{selfdestruct} functions and 6 reasons why they do not add them to their smart contracts. 
	
	\item We design an approach to find 5 reasons why smart contracts self-destructed. These self-destruct reasons can be used as a guidance when practitioners develop their contracts. Also, our approach gives inspiration for researchers. They can use the same approach to find more self-destruct reasons and apply the method to other smart contract platforms, e.g., Ethereum Classic\footnote{Ethereum Classic is another popular blockchain platform which support the running of smart contracts.}~\cite{ETC}.
	
	\item We propose a tool named \textsc{LifeScope} to detect two problems that might shorten the life span of smart contracts.  \textsc{LifeScope} obtains 100\% of F-measure in detecting \textit{Unmatched ERC20 Token}. And it achieves an F-measure and AUC of 77.89\% and 0.8673, respectively in detecting \textit{Limits of Permission}. 
	
\item According to the feedback from our survey, there are six common reasons why some developers do not use \textit{selfdestruct} function. We give five suggestions for developers to address these issues and to help them better use the\textit{selfdestruct} function in their smart contacts. 
\end{itemize}		

The organization of the rest of this paper is as follows. In Section 2, we present the background knowledge of smart contracts. Then, we show the answer to the three research questions in Section 3-5, respectively. We discuss the implication, how to better utilize \textit{selfdestruct} function and threats to validity in Section 6. After that, we introduce related works in Section 7. In Section 8, we conclude the whole work and present our future work.

	\section{Background}
\label{background}
In this section, we briefly introduce the background information about smart contracts, the Ethereum system, and some features and knowledge about smart contract programming. \vspace*{-0.1cm}

\subsection{Smart Contracts } 

Bitcoin was the first cryptocurrency that utilized blockchain as its underlying technology. It allows users to encode scripts to process transactions. However, the scripts on Bitcoin are not Turing-complete, which restricts the usage of Bitcoin~\cite{bitcoin}. In contrast, Ethereum leverages a technology named \textit{smart contracts}. These can be regarded as self-executed programs that run on the blockchain. When developers deploy smart contracts to Ethereum, the source code of the contracts will be compiled into bytecode and reside on the blockchain forever. The storage of Ethereum is very expensive, as all the data stored on the blockchain will be copied on each node, a so-called distributed ledger. To minimize the data space, the source code of the smart contracts will not be stored on the blockchain. Once a contract is deployed to the blockchain,  the contract is identified by a 20-byte hexadecimal address. Arbitrary users can call the functions of a smart contract by sending transactions to the contract address.

\begin{lstlisting}[caption={A simple contract},label=List:example]
pragma solidity ^0.4.25;
contract Example{
  address owner_addr;
  address[] participators;
  uint participatorID = 0;
  function constructor(){
	owner_addr = msg.sender; 
  }
  function() payable{
	if(msg.value != 1 Ether)
	  revert();
	participators[participatorID] = msg.sender;
	participatorID++;
	if(this.balance == 10 Ether)
		getWinner();
  }
  function getWinner(){
	uint random = uint(block.blockhash(block.number)) % participants.length;
	participators[random].transfer(9 Ether);
	participatorID = 0;
  }
  modifier onlyOwner{ 
	if(msg.sender != owner_addr) 
	  _;
  }
  function Selfdestructs(address addr) onlyOwner(){
	selfdestruct(addr);
  }
}

\end{lstlisting} 

Listing~\ref{List:example} is an example of a smart contract that implements a simple gambling game by using Solidity~\cite{Solidity}. Solidity is the most popular smart contract programming language on the Ethereum platform. Users can send 1 Ether to the contract. Once the contract receives 10 Ethers, the contract will choose 1 user as the winner randomly and send 9 Ethers  to him/her. 

The first line is called the version pragma, which is used to identify the compiler version of the contract. Lines 3-5 are the global parameters, and the function on line 6 is the constructor function of the smart contract.  The constructor function can only be executed once when deploying the contract to the blockchain. Therefore, this function is usually used to store the owner's information. Specifically, line 7 stores the owner's address by using \textit{msg.sender}. (\textit{msg.sender} is used to obtain the address of the transaction sender. ) Function on line 9 named fallback function, which is the only unnamed function of the smart contract. This function will be executed automatically when an error function call happens. For example, a user calls function ``$\delta$", but there is no function named  ``$\delta$" in the contract. In this situation, a fallback function will be executed to handle the error call. If the fallback function is marked by a keyword named \textit{payable}, the fallback function will also be executed automatically when the contract receives Ethers. Lines 10 and 11 guarantee that each user sends 1 Ether to the contract. If the user sends other amounts of Ethers, the transaction will be rolled backed by executing \textit{revert()}, which is a function provided by \textit{Solidity}. When the contract receives 10 Ethers (line 14), the contract will choose 1 user to send 9 Ethers by using function \textit{getWinner} (line 17). The contract generates a random number by using the block info related functions\footnote{ (block.blockhash and block.number are the functions provide by Solidity to obtain block related information. Since block hash number is random; so it can be used to generate random numbers sometimes. ) } in line 18. Then, the contract sends 9 Ether to the winner in line 19.  

\subsection{Function Modifier}

Ethereum is a permission-less network -- everyone can call methods to execute smart contracts. Developers usually add permission checks for permission-sensitive functions.  For example, the contract in Listing~\ref{List:example} records the owner's address in its constructor function (line 7). In this case, the contract can compare whether the caller's address is the same as the owner's address. Solidity provides \textit{Function Modifiers} which are used to add prerequisites checks to a function call. A function with function modifier can be executed only if it passes the check of the modifier.

Listing~\ref{List:example} line 22 shows a modifier named \textit{onlyOwner}. This modifier requires the transaction creator (\textit{msg.sender}) should be the owner of the contracts (\textit{owner\_addr}). Function \textit{Selfdestructs} on line 26 contains this modifier. Therefore, only the owner of the contract can call \textit{selfdestruct} function in line 27. 

\subsection{Selfdestruct Function }
\label{sec:back_selfdesctuct}
The \textit{selfdestruct} function in Listing 1 line 27 is the only way to remove the contract from Ethereum.  When executing this method, the caller can transfer all Ethers on balance to a specific address (\textit{addr}) (line 27). Then, the contract will be discarded. If others transfer Ethers to the self-destructed contract address, the Ethers will be locked forever. Calling \textit{selfdestruct} function when a contract is no longer needed can help clean up the Ethereum environment. To motivate this, Ethereum refunds up to half of the gas used by a contract transaction calling the \textit{selfdestruct} function to the transaction sender. This mechanism is also utilized by GasToken~\cite{gastoken}, which allows users to store gas when the gas price is low and use the gas when it is expensive. Specifically, a user can create a simple contract (GasToken) that contains \textit{selfdestruct} function when the gas price is low. Then, the user can destruct the GasToken to save the gas when the gas price is high. However, GasTokens also have the downside to the Ethereum network, as it leads to the creation of millions of ``useless" contracts, which is against the original motivation of the gas refund. EIP-3529~\cite{EIP-3529} and EIP-3298~\cite{EIP-3298} are two Ethereum improvement proposals that suggest reducing the gas refunds.  EIP-3529 recommends reducing the gas refund from up to 1/2 gas used by a transaction to 1/5, and EIP-3298 even recommends removing the gas refund directly. These two EIPs imply that the GasToken might be nullified in the future.

The \textit{selfdestruct} function is sometimes harmful as the immutability feature can be broken. Immutability is a special and important feature of smart contracts compared to traditional programs. Once a contract is deployed to the blockchain, none can modify the contract, even the owner. However, this function can allow the owner to kill the contract and make the contract disappear from the blockchain. This might reduce the confidence of the users, as the owner can transfer all the Ethers of the contract. For example, the owner can transfer all the Ethers by calling the \textit{selfdestruct} function on contract in Listing~\ref{List:example} when the contract receives 9 Ethers. In this case, all the users are losers.

\begin{lstlisting}[caption={The Demo of the DAO Attack}, label=List:Reentrancy]
contract Victim {
	mapping(address => uint) public userBalannce;
	function withDraw(){
		uint amount = userBalannce[msg.sender];
		if(amount > 0){
			msg.sender.call.value(amount)();
			userBalannce[msg.sender] = 0;
		}
	} 
		...
}
contract Attacker{
	function() payable{
		Victim(msg.sender).withDraw();
	}
	function reentrancy(address addr){
		Victim(addr).withDraw();
	} 
		...
}
\end{lstlisting}

\subsection{The DAO Attack - A Motivation Example of the \textit{selfdestruct} Function}
In 2016, attackers found a vulnerability named Reentrancy~\cite{oyente, smellDefinition} in a smart contract of the Decentralized Autonomous Organization (DAO organization), and this vulnerability made the DAO organization lost 3.6 million Ethers (\$270/Ether on Feb. 2020). People usually call this infamous attack a DAO attack.

List~\ref{List:Reentrancy} is a demo of the DAO attack. There are two smart contracts, i.e., \textit{Victim} contract and \textit{Attacker} contract. The \textit{Attacker} contract is used to transfer Ethers from \textit{Victim} contract, and the \textit{Victim} contract can be regarded as a bank, which stores the Ethers of users. Users can withdraw their Ethers by invoking \textit{withDraw()} function. However, \textit{withDraw()} function contains the \textit{Reentrancy} vulnerability in line 6-7. 

First, the \textit{Attacker} contract uses \textit{reentrancy()} function (line 16) to invokes \textit{Victim} contract’s \textit{withDraw()} function in line 3. The \textit{addr} in line 17 is the address of the \textit{Victim} contract. Normally, the \textit{Victim} contract  sends Ethers to the callee in line 6, and resets callee's balance to 0 in line 7. However, Ethereum does not support concurrency, which means \textit{Victim} contract sends Ethers to Attacker contract before resetting the balance to 0. When the \textit{Victim} contract sends Ethers to the \textit{Attacker} contract, the fallback function (line 13) of the \textit{Attacker} contract will be invoked automatically, and line 7 is not executed at that time. So, the \textit{Attacker} contract can invoke \textit{withDraw()} function repeatably.

Actually, the DAO attack continued for several days and the organization even noticed that their contract had been attacked at that time. However, they could not stop the attack or transfer the Ethers because of the immutability feature of smart contracts. If the contract contains a \textit{selfdestruct} function, the DAO organization can transfer all the Ethers easily, and reduce the financial loss. 

%\subsection{Predecessor Contracts and Successor Contracts}
%Using the \textit{selfdestruct} function is the only way to remove code from the blockchain. After executing this function, the contract will be destroyed, and there is no way to recover it. In this paper, we call a smart contract that has executed a \textit{selfdestruct} function a \textit{'self-destruct'} contract. There are many reasons why contracts might be designed to self-destruct, e.g., to mitigate a later found security vulnerability or to enable a later functional change. Many developers will upgrade the \textit{self-destruct contract} and deploy a new version. We call the new version of the contract a \textit{'Successor'} Contract, and the self-destructed contract is the \textit{Predecessor} contract of the \textit{Successor} contract.

\subsection{ERC20 Standard}
\label{sec:erc20}
Motivated by the great success of Bitcoin, thousands of cryptocurrencies have been created in recent years. However, most of them do not have their own blockchain system. Instead, they are usually implemented by smart contracts that run on the Ethereum, also called \textit{tokens}. To ensure different tokens can interact accurately and be reused by other applications (e.g., wallets and exchange markets), Ethereum provides ways to standardize their behaviors. ERC20~\cite{erc20} is the most popular token standard on Ethereum. It defines 9 standard interfaces (3 are optional) and 2 standard events. To design ERC20 tokens, developers should strictly follow the standard. For example, the standard method \textit{transfer} is declared as ``\textit{function transfer(address \_to, uint256 \_value) public returns (bool success)}''. This function is used for transferring tokens to a specific address (\textit{\_to}). The ERC20 standard requires this function to throw an exception if the caller's account balance does not have enough tokens to spend. Besides, the function should fire an event named ``\textit{TRANSFER}'' to inform the caller whether the tokens are transferred successfully.

\subsection{Card Sorting}

Card sorting is a research method to organize data into logical groups~\cite{spencer2009card}. Due to the low-tech and inexpensive nature of card sorting, it is widely used to help users understand how users would organize and structure the data that makes sense to them. The users who conduct a card sorting process first need to identify the key concepts and write them into labeled cards, which can be actual cards or a piece of paper. Then, they are asked to classify them into groups that they think are appropriate. By utilizing card sorting, users can design workflow, architecture, category tree, or folksonomy.

There are three kinds of card sorting, i.e., open card sorting, closed card sorting, and hybrid card sorting. Open card sorting is commonly used for organizing data with no predefined groups. Specifically, each card will be clustered into a group with a certain topic or meaning first. If there is no appropriate group, a new group will be generated. All the groups are low-level subcategories and will be evolved into high-level subcategories further. Closed card sorting is used for organizing data with predefined groups. Each card is required to cluster into one of the groups. Hybrid card sorting combines open card sorting and closed card sorting. Hybrid card sorting has predefined groups but allows the creation of new groups during the process.

	%\newpage
\section{RQ1: Developer's Perspective about selfdestruct Function}
\label{sec:RQ1}
\subsection{Motivation}
Usage of the \textit{selfdestruct} function can enable developers to destruct their contracts and transfer Ethers when emergency situations happen, e.g. a contract is being attacked or is found to be buggy. However, this function is also harmful for both contract users and contract owners. In our analysis, we crawled all of the 54,739 verified smart contracts from Etherscan~\cite{EtherScan} by the time of writing (for details see Section~\ref{sec:dataset}), and found 2,786 (5.1\%) smart contracts contain a \textit{selfdestruct} function in their source code. In this RQ, we aim to investigate the developers' perspective about using the \textit{selfdestruct} function in Ethereum smart contracts. By understanding the reason why they include or exclude \textit{selfdestruct} functions in their contracts, we can better understand the advantages and disadvantages of this function. We then want to design some guidance about using the \textit{selfdestruct} function (can be found at Section~\ref{sec:suggestion}), which enables developers to design a more robust smart contract.

\subsection{Approach} 
\subsubsection{Validation Survey}
%\noindent{ \textbf{3.2.1. Validation Survey: }} 
In this paper, we utilize the methods proposed by Kitchenham et al. ~\cite{kitchenham2008personal} to design a survey for collecting the opinions from smart contract developers. To increase the response rate, we make the survey anonymous~\cite{tyagi1989effects} and provided a raffle for developers who take part in our survey. Participation in the raffle is voluntary; we chose two respondents who provided their email addresses as the winner, and gave them \$50 Amazon gift cards as the reward. We first use a small scale survey to collect feedback about our survey. The feedback includes: (1). Whether the expression about our question is easy to understand. (2). Whether the time to finish the survey is reasonable. After the small scale survey, we refine our questionnaire based on the feedback we collected. Finally, conduct a large scale investigation to collect our data. \footnote{ETHICS COMMITTEE APPROVAL}

\subsubsection{Survey Design}
%\noindent{ \textbf{3.2.2. Survey Design: }} 
To understand the background of the respondents better, we first collect their demographic information. These five questions can help us have an overall understanding of the respondents. 

\textit{\textbf{a. Demographics: } }

\begin{itemize}\setlength{\itemsep}{1pt}
	\item 1. Professional smart contract developer? : Yes / No
	
	\item 2. Involved in open source software development? : Yes / No
	
	\item 3. Main role in developing smart contract: Testing / Development / Management / Other
	
	\item 4. Experience in years (decimals ok)? 
	
	\item 5. Current country of residence ? 
	
\end{itemize}

After that, the respondents are required to choose yes / no in the question 6. If they choose yes, they are required to answer question 7; otherwise, they should answer question 8. Both question 7 and 8 contain a textbox, which enables respondents to input their answer. (There is no length requirement / restriction of their inputs.)

\textit{\textbf{b. Questions about selfdestruct Functions: }}

\begin{itemize}\setlength{\itemsep}{1pt} 
	\item 6. Will you add \textit{selfdestruct} functions in your future smart contracts? : Yes (Go to Q7) / No (Go to Q8)
	
	\item 7. Why do you add \textit{selfdestruct} functions? 
	
	\item 8. Why do you not add \textit{selfdestruct} functions? 
	
	%	\item 9. If you have a smart contract which contains \textit{selfdestruct} function, in which situations you will call it?
\end{itemize}

To increase the response rate, we prepare two kinds of survey\footnote{The two surveys and related feedback can be found at: https://github.com/Jiachi-Chen/Selfdestruct}, i.e., English Version and Chinese Version, as Chinese is the most spoken language and English is an international language in the world. The Chinese version survey is carefully translated to ensure the contents between the two versions are the same.  

\subsubsection{Recruitment of Respondents}
%\noindent{ \textbf{3.2.3. Recruitment of Respondents: }}
To receive sufficient response from different backgrounds, we first sent our questionnaire to our contacts who are working in world-famous blockchain companies or doing related research in academic institutions, e.g., \textit{Ant Financial, The Hong Kong Polytechnic University, NUS, The University of Manchester}. Then, we also collect developers' email addresses on their Github homepage who are contributing to open-sourced blockchain projects. 
We collected 1,238 email addresses from Github. Due to the scale of the smart contract projects, 1,238 are the numbers of contributors of the top 100 most popular (ranked by stars) smart contract related projects, which is a good number compared to previous smart contract related surveys~\cite{chen2020defining, bosu2019understanding, chakraborty2018understanding} . 

\subsection{Result}
Since some email addresses we collected are illegal or abandoned, we successfully sent our survey to 996 developers, and receive 88 responses from 32 countries (The response rate is 8.84\%). The top three countries in which respondents reside are China (29.89\%), the USA (8.05\%) and the UK (5.57\%). Three of the respondents claim that they are not professional smart contract developers and have no experience in developing smart contracts. Therefore, we exclude their responses and use the remaining 85 responses for analysis. 
The average years of experience in developing Ethereum smart contracts are 1.96 years (standard deviation is 1.05) for all of our respondents. As the survey was undertaken in Sept. 2019 (about 4 years since Ethereum was first published), 1.96 average years of experience shows that they have good experience in developing smart contracts. Among these respondents, 62 (72.94\%), 10 (11.76\%), 5 (5.88\%) described their job roles as development, testing, and management, respectively. The other 8 respondents said they have multiple roles, such as security auditor and research.

Guided by previous works~\cite{zou2019smart, chen2020defining}, we performed open card sorting~\cite{spencer2009card} to analyze the survey feedback to summarize the reasons why developers include or exclude the \textit{selfdestruct} function in their smart contracts. Feedback that we received in Chinese was first translated into English. After that, we manually converted each feedback into several separate units with coherent meaning, as some feedbacks contain several reasons. Then, a card was created for each separate unit with a title (ID) and description (feedback content). Two experienced smart contract researchers were involved in the card sorting. The detailed steps are: 

\textit{Iteration 1: } Two researchers randomly chose 20\% of the cards. Each card was analyzed by the two researchers together. They were required to summarize a detailed reason, e.g, Security concern, Trust concern. If the root concern is unclear, they omit the card from the card sort.

\textit{Iteration 2: } The same two researchers analyzed the remaining 80\% of the cards independently by following a similar method as iteration 1.  Some new reasons are found in this step. After they have gone through the cards independently, they compared their results and discussed any differences. Finally, 6 reasons for including and 6 reasons for excluding \textit{selfdestruct} function were summarized. 

There is a threat that what developers told us, i.e., their reasons for adding a self-destruct function to their smart contracts, may be different from what they do in reality. We do not claim that our user study is final and complete; rather, we view it as a first step to better understand the usage of \textit{selfdestruct} function. We invite others to replicate our study with additional surveys and interviews. 

\subsubsection{Reasons for including the \textit{selfdestruct} function}
%\noindent  \textbf{3.3.1. Reasons for including the \textit{selfdestruct} function: }
33 (38.82\%) of the respondents claim that they will add the \textit{selfdestruct} function in their smart contracts. We analyzed the feedback of these respondents and summarized five key reasons. \textbf{As some respondents give more than one reason, the sum of these is higher than 33. }

%\textbf{Reasons: }
%\begin{itemize}\setlength{\itemsep}{1pt}

%	\item 1. Security concern. 18

%	\item 2. Clean up the blockchain environment.  11

%	\item 3. Quickly withdraw Ethers.  9

%	\item 4. For upgrading contracts. 4

%	\item 5. Business Requirement. 2
%\end{itemize}		

%	It is clear that the most common reason for adding \textit{selfdestruct} function  is for security concern. 18 of the respondents claim that they can use this function to stop the contracts when vulnerabilities are detected in their contracts. After fixing the vulnerabilities, they can deploy a new contract. The second reason is for cleaning up the blockchain environment. 11 of respondents mention that the contracts have their life circle. When their duty is finished, they will call \textit{selfdestruct} function to remove them from the blockchain. Therefore, no-one is able to call the contract, and it can free up chain space. 9 of the respondents told that this function allows them to withdraw the Ethers quickly. 4 respondents said they will use the function to upgrade their contracts in the future. After deploying a new contract, the old version should be removed. Other 2 respondents said their business parters require them to add \textit{selfdestruct} function. 

\, \textbf{\textit{Reason 1: Security Concerns.} }  \textbf{18 (54.55\%) respondents} claim that they  use the \textit{selfdestruct} function to stop the contracts when security vulnerabilities are detected in their contracts. After fixing the vulnerabilities, they can deploy a new contract.

\, \textbf{\textit{Reason 2: Clean Up Environment.}}  Blockchain is a distributed ledger where each node stores all the data. After destructing the contracts, the functions of the contracts cannot be called anymore.  \textbf{11 (33.33\%) respondents} mention that when the duty of the contract  is finished, they will call the \textit{selfdestruct} function to remove the contracts from the blockchain, which can clean up the blockchain environment. 

\, \textbf{\textit{Reason 3: Quickly Withdraw Ethers.} } By using the \textit{selfdestruct} function, the owner of the contract can remove all the Ethers to a specific address. \textbf{9 (27.27\%) respondents} claim this function can help them transfer assets quickly. 

\, \textbf{\textit{Reason 4: Upgrade Contracts.} }  \textbf{4 (12.12\%) respondents} said they may need to upgrade their contracts in the future. Adding a \textit{selfdestruct} function is the easiest method to upgrade their contract. This function allows them to remove the old version of the contract and deploy a new version.  %Although using ``DelegateCall" can also implement upgradeable contract, using \textit{selfdestruct} function can reduce the cost and difficulty of development.

\, \textbf{\textit{Reason 5: Business Requirement.} } The business requirement is also a reason why developers add \textit{selfdestruct} function. \textbf{2 (6.06\%) respondents} said their business partners require them to add the \textit{selfdestruct} function. 

\, \textbf{\textit{Reason 6: Gas Refund.} } As we introduced in Section~\ref{sec:back_selfdesctuct}, the gas refund feature of the \textit{selfdestruct} function allows the transaction sender to get up to half of the gas back. \textbf{1 (3.03\%) respondent} mentioned that he/she adds the \textit{selfdestruct} function to get the gas back. 

According to our survey, 22 / 33 respondents claim that they add \textit{selfdestruct} function for security concerns or to upgrade contracts. These two motivations can lead to redeployment of smart contracts after developers destruct the contracts.  Besides, the survey feedback also show that \textit{selfdestruct} function is useful for contract developers to handle emergency situations, e.g., when serious security issues are found in the contracts.

\subsubsection{Reasons for excluding the \textit{selfdestruct} function}
%\noindent \textbf{3.3.2. Reasons for excluding the \textit{selfdestruct} function: }
52 (61.18\%) of the respondents claim that they will not add \textit{selfdestruct} functions in their smart contracts.  \textbf{As some respondents give more than one reason, the sum of these is higher than 52. }

% \textbf{Reasons: }
%\begin{itemize}\setlength{\itemsep}{1pt}

%	\item 1. Security concern. 19

%	\item 2. Trust concern. 16

%	\item 3. No needs. 16

%	\item 4. Do not familiar with it. 7

%	\item 5. Additional complexity. 4

%	\item 6. Risk of Ether losing after destruct contract. 2

%\end{itemize}		

%There are six reasons why they exclude \textit{selfdestruct} function from their smart contracts. The \textit{security concern} is the most important reason. 19 of the respondents worried that \textit{selfdestruct} function might open an attack vector for adversaries to exploit. Trust concern is the second most popular reason. People who give this reason believe the immutability is an important feature for smart contracts. However, including a \textit{selfdestruct} function might increase their truthlessness. 16 of respondents mention that their contracts do not need this function  as their contracts do not have Ethers or their developmental requirements do not contain this requirement. When they want to upgrade the contract, they just discard the old contract. 7 of respondents claim that they do not familiar with \textit{selfdestruct} function. So, they do not add it. 4 people told that they need to add more test if they add \textit{selfdestruct} function in the contracts, which can add the additional complexity of their contracts. Risk of Ether losing after destructing the contract is also a big concern for the developers. 2 developers worried that people may sent Ethers to the self-destruct contract and these Ethers will be locked forever. 

\, \textbf{\textit{Reason 1: Security Concerns.}} The \textit{selfdestruct} function can also lead to serious security problems if the contract does not handle access permissions correctly ﻿or the private keys of owners are leaked. \textbf{19 (36.54\%) respondents} worried that the \textit{selfdestruct} function might open an attack vector for adversaries to exploit. Limiting the permission of calling selfdestruct function is not difficult. For example, a contract can only allow specific addresses to execute this function. However, it is also possible that the private keys~\footnote{There are two kinds of accounts in Ethereum, i.e., externally owned account (EOA) and contract account. A contract account is controlled by its code, and an EOA is controlled by the private key.} of these addresses might be stolen. Once the private keys are stolen by attackers, the smart contract can then be destructed by attackers and all the Ethers will be lost.

\, \textbf{\textit{Reason 2: Trust Concerns.} } \textbf{16 (30.77\%) respondents} who give this reason believe that including a \textit{selfdestruct} function might lead to trust concerns from the contract users.  To be specific, a \textit{selfdestruct} function allows the owner to kill the contract and make the contract disappear from the blockchain. Also, the owner can transfer all the Ethers, which raises a trust concern for the user.

\, \textbf{\textit{Reason 3: Requirement Concerns.} } \textbf{16 (30.77\%) respondents} mention that their contracts do not use \textit{selfdestruct}  function as their contracts do not have Ethers. Therefore, they do not need to transfer Ethers. Besides, when they want to add some new functionalities, they said they could deploy a new smart contract and ignore the old one. 

\, \textbf{\textit{Reason 4: Unfamiliarity.}}  \textbf{7 (13.46\%) respondents} claim that they are unfamiliar with \textit{selfdestruct} function. They are worried that they might misuse the \textit{selfdestruct} function, and lead to the bugs. 

\, \textbf{\textit{Reason 5: Additional Complexity.}} \textbf{4 (7.69\%) respondents} told us that they need to add more tests if they add \textit{selfdestruct} functions in the contracts, which can introduce additional complexity to their contracts. 

\, \textbf{\textit{Reason 6: Additional Financial Risk.}} Risk of losing Ether after destroying the contract is also a concern for the developers. \textbf{2 (3.85\%) respondents} worried that people may send Ethers to the self-destructed contract, and these Ethers will be locked forever. 

%According to our survey feedback, we find that the \textit{selfdestruct} function might be risky for both contract developers and contract users. For contract developers, they need to pay more effort to ensure only specific people have permission to call the function, which might increase the cost of development. Besides, how to ensure no one will send Ethers to the self-destructed contract is also a big question. For contract users, their balance might be stolen by the contract owners, as the owners have the ability to transfer all the Ethers. 

%The survey is also the motivation for the next two RQs. Specifically, from the survey, we find that there are five reasons why developers destruct their contracts, and only the reason “Security concern” and “Upgrade contracts” can lead to the redeployment of smart contracts. These two reasons are also the most common reasons (66.67%) why  developers destruct their contracts. This finding gives us the motivation of RQ2 i.e. that we can find some security issues by comparing two versions of contracts.  Without RQ1, there is no evidence to prove that developers will use the selfdestruct function to destruct the contract and deploy a new version when security issues are found. 

According to our survey feedback, \textit{selfdestruct} function might be risky for both contract developers and contract users. We find six reasons why developers destruct their contracts, and the reasons ``Security concern" and ``Upgrade contracts” can lead to the redeployment of smart contracts. These two reasons are also the most common reasons (66.67\%) why  developers destruct their contracts. This finding gives us the motivation of RQ2 that we can find some security issues by comparing two versions of contracts.

	\section{RQ2: Reasons for Self-destruct}
\label{sec:RQ2}
\subsection{Motivation}
\label{sec:RQ2-Motivation}

According to our survey, 22 out of 33 respondents claim that they add a \textit{selfdestruct} function for security concerns or to upgrade contracts. These two motivations can lead to redeployment of smart contracts after developers destruct the contracts. Therefore, by comparing the difference between the two versions of the contract, we can find the reasons why contracts self-destructed. Consider the following scenario.

Bob is a smart contract developer. He developed a smart contract several weeks ago, and his company uses this contract to receive money from other companies. However, they find that a function in the smart contract does not limit the caller's permission, which can lead to serious security problems. Therefore, Bob has to destroy the contract by involving \textit{selfdestruct} function and deploy a new contract to the blockchain. The new contract adds a permission check to avoid this vulnerability. Bob and his colleagues try their best to inform other companies not to transfer Ethers to the self-destructed contract anymore. However, it requires a long time to inform all companies. Many users still transfer Ethers to the self-destructed contract, and all the Ethers send to the contract are lost forever. It causes a great financial loss to Bob's company. 

From this scenario, we see that calling the \textit{selfdestruct} function may lead to great financial loss. Therefore, we should try to make contracts robust. If we tell Bob that many previous contracts are destructed because a function in the smart contract does not limit the caller’s permission, he might check whether his contract contains the same problem and can avoid this problem.  

In this section, we compare the self-destructed contracts and their successor contracts to summarize reasons why contracts self-destructed. The reasons we identify can guide smart contract developers and help them refine their contracts.  

\subsection{Approach}

\begin{figure}
	\begin{center}
		\includegraphics[width=0.79\textwidth]{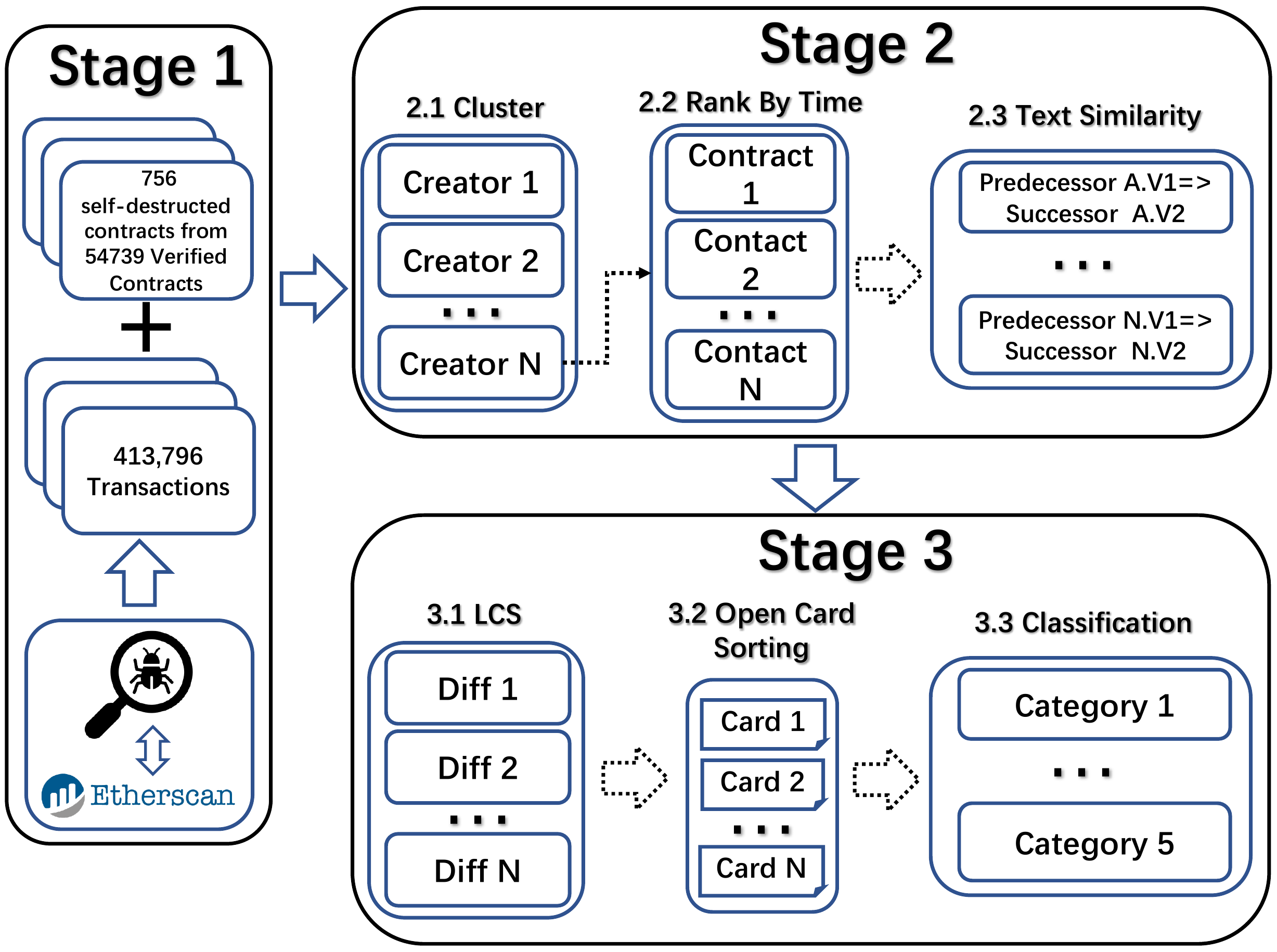} 
		\caption {Overview architecture of finding the self-destructed reasons} 
		\label{Fig:suicdeReason}
	\end{center}
\end{figure}

\begin{table}
	\footnotesize
	\caption{Information for the 756 self-destructed contracts }
	\label{tab:info_contracts}
	\centering
	%\begin{tabular}{c | c}
	\begin{tabular}{c | c | c | c | c }
		\hline
		 & Max & Min & Avg. & Median\\
		\hline
		Life Span & 879.6 days & $\leq$ 1 hour & 40.8 days & 4.3 days \\
		\hline
		No. Trans & 100582 & 2 & 538.8 & 7 \\
		\hline 
		No. Eths & 208.6 & 0 & 0.44 & 0 \\
	\end{tabular}	
\end{table}

Figure~\ref{Fig:suicdeReason} depicts the detailed steps to identify the reasons why some smart contracts have been destructed. Our method consists of three stages. In the first stage, we crawl all verified contracts and their transactions from Etherscan. We crawled 54, 739 smart contracts altogether. We found that 2,786 (5.1\%) of these smart contracts among 54,739 contracts contain a \textit{selfdestruct} function, and 756 (27.14\%) contracts have been destructed. In the second stage, we first divide crawled contracts into several groups by their creators' addresses. In this case, we can find smart contracts that are created by the same authors. We only use groups that contain self-destructed contracts and rank all the contracts in the same group by contracts' creation time. Then, we compute the code similarity of contracts in each group to find self-destructed contracts (also called predecessor contracts) and their successor contracts. In the last stage, we compare the difference of predecessor contracts and their successor contracts by using \textit{open card sorting}. Finally, we summarized 5 reasons why contracts self-destructed. 

\subsubsection{Stage 1. Data Collection:}
\label{sec:dataset}
\begin{figure}
	\begin{center}
		\includegraphics[width=0.85\textwidth, height=0.55\textheight]{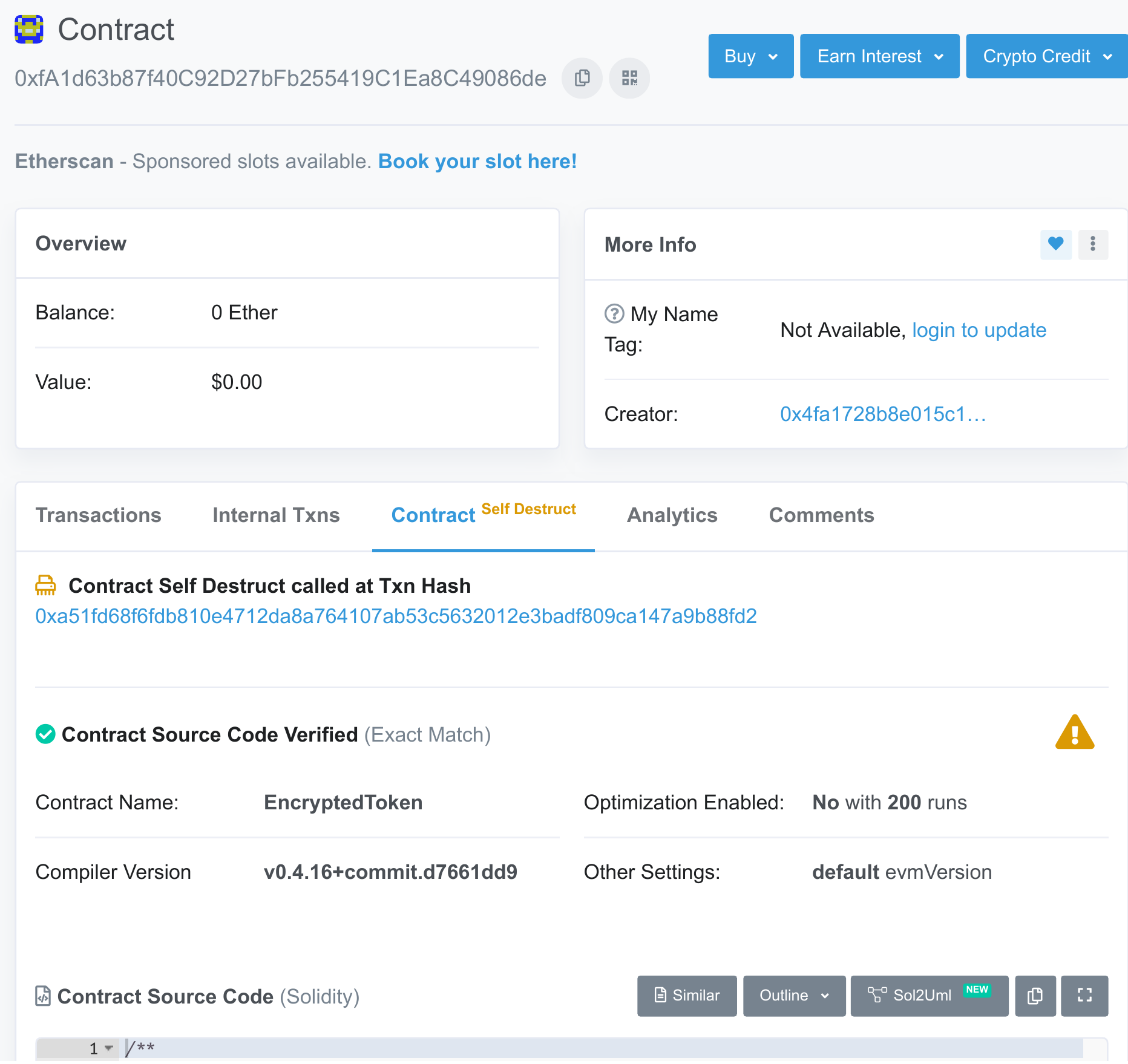} 
		\caption {A smart contract on Etherscan} 
		\label{Fig:contract}
	\end{center}
\end{figure}

%\noindent{ \textbf{4.2.1 \, Stage 1: Data Collection}}
Stage 1 is used to collect data for the following two stages. Our data contains three parts, i.e., \textit{verified contracts}, \textit{self-destructed contracts}, and \textit{contract transactions}.

\ \noindent \textbf{Verified Contracts}: Verified contracts are crawled from Etherscan. To crawl the source code of verified smart contracts from Etherscan, we first need to know the contract addresses of verified contracts. Figure~\ref{Fig:contract} is a smart contract on Etherscan. By obtaining the contract address, we can easily download the source code, transactions, and other information of the contract. The contract address list is provided by Etherscan (https://etherscan.io/contractsVerified). However, Etherscan only shows the last 500 verified contract addresses since Jan. 2019. The data used in this paper are crawled before Jan. 2019 and the lasted 500 verified contracts by the time of the writing. We finally obtained 54, 739 verified contract addresses. After obtaining the contract addresses, we can crawl the source code from Etherscan directly. 

\ \noindent \textbf{Self-destructed Contracts}:  Finding whether a verified smart contract has been self-destructed is straightforward. If a contract has self-destructed, there will be a label (\textit{Self Destruct}) given on Etherscan (see Figure~\ref{Fig:contract}). We found 756 self-destructed contracts from 54,739 verified smart contracts. The detailed information (creation time, destructed time, number of transactions / balances ) of each contract can be found at https://zenodo.org/record/5518527\#.YUl-bWYzYUE/Contract\_Info\_Suicide.csv. Table~\ref{tab:info_contracts} shows a brief summary of these 756 contracts. Life span is the time interval between the transaction of creation and destruction. No. Trans / Eths records the transactions and balance of a contract before its destruction, respectively.

\begin{figure}
	\begin{center}
		\includegraphics[width=0.95\textwidth]{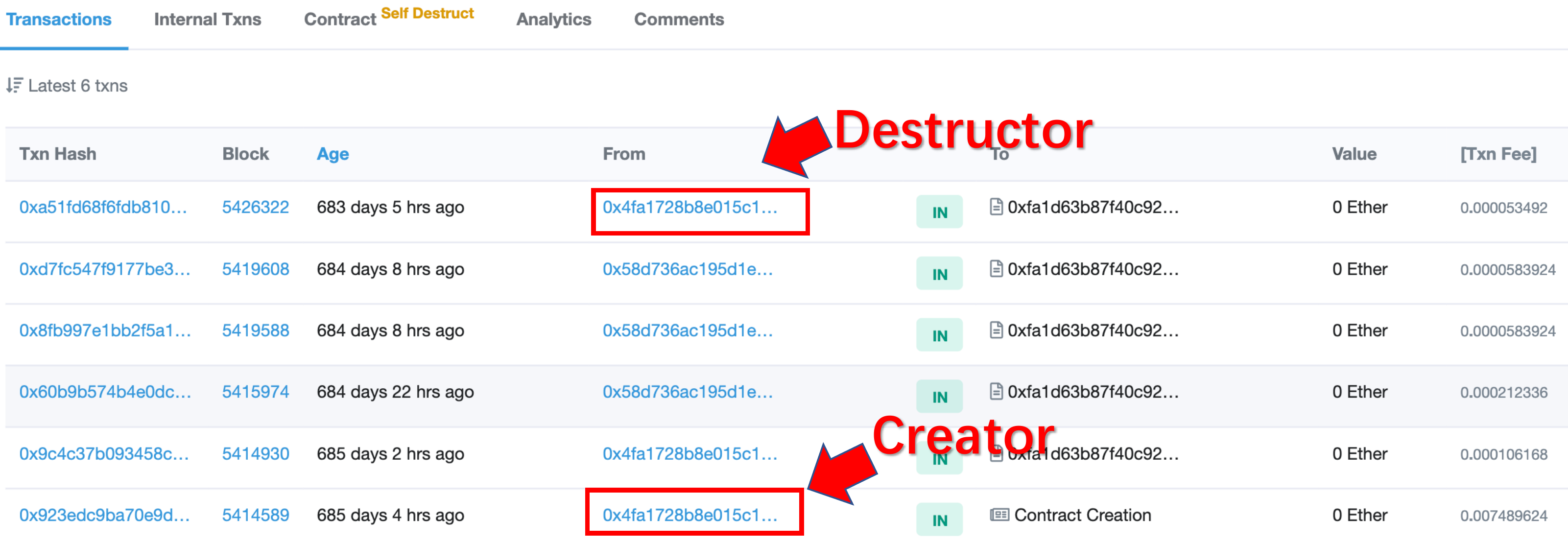} 
		\caption {Transactions of a Self-destructed Contract} 
		\label{Fig:trans}
	\end{center}
\end{figure} 

\ \noindent \textbf{Transactions}: Transactions on Ethereum record the information of the external world interacting with the Ethereum network. All the transactions can be found on Etherscan. We collect all 413,796 transactions of 756 self-destructed smart contracts. Figure~\ref{Fig:trans} is the transactions of a self-destructed contract. In the first transaction, we can find who deployed the contract (creator) and we can find who destructed the contract (destructor) in the last transaction.

\subsubsection{Stage 2. PS Pairs Generation:}
%\noindent{ \textbf{4.2.2 \, Stage 2: Pairs Generation}}
A \textit{PS} (predecessor and successor) pair is denoted as: $PS = \langle \mathbb{P}, \mathbb{S} \rangle$, where $\mathbb{P}$ is a smart contract that has executed the selfdestruct function, named as the predecessor contract. $\mathbb{S}$ represents $\mathbb{P}$'s upgradeable version, that we call a successor contract.  $\mathbb{P}$ and $\mathbb{S}$ are deployed by the same address and have similar functionalities.

The aim of stage 2 is to find all the PS pairs in our dataset, which helps to reduce the manual effort when summarizing the reasons for self-destruct in stage 3.
For some self-destructed contracts, it is not easy to find their successor contracts. For example, in our survey feedback, some developers mentioned that they added the \textit{selfdestruct} function is only to quickly transfer Ethers and will not redeploy the contract to Ethereum. For this kind of contracts, we cannot find their successor ones, and thus we removed them from our analysis list. Some developers said they would destruct the contracts when bugs are found. Then, they will redeploy a smart contract after fixing the bug. Thus, we can compare the difference of two versions of contracts to find the bugs, and the two versions of the contracts are likely to have high similarity. In this stage,  we calculate the similarity to find the successor contracts of the self-destructed contracts. We found 436 contracts among 756 self-destructed contracts that have successor contracts.

\textbf{Step 2.1 Cluster}: We first find the creator addresses of all the 54,739 verified smart contracts through their transactions. In this step, we inspect the first transaction of each smart contract as the first transaction contains the creator address and creation time. Then, we classify the contracts into several groups according to their creator addresses. If two contracts have the same creator address, they will be classified into the same group. We only choose groups that contain self-destructed contracts. 

\textbf{Step 2.2 Rank by Time}: A \textit{PL} (predecessor and alive) pair is denoted as: $PL = \langle \mathbb{P}, \mathbb{L} \rangle$, where $\mathbb{P}$ is a smart contract that has executed the selfdestruct function. $\mathbb{L}$ represents a smart contract that has been deployed later than $\mathbb{P}$. $\mathbb{P}$ and $\mathbb{L}$ are deployed by the same address. 

In this step, we first rank contracts in each group by their creation time, which can be obtained from the first transaction. Then, we can obtain several \textit{PL} pairs. For example, one group contains five contracts, they are contract \textit{a,b,c,d,e} and these five contracts are ranked by creation time. Contract b and d are the self-destructed contacts in these five contracts. Finally, we output four PL pairs, i.e., \textit{(b,c), (b,d), (b,e)} and \textit{(d,e)}. 

\textbf{Step 2.3 Text Similarity}:  We compute the code similarity between two contracts to identify whether the later created contract is the successor contract of the self-destructed contract. \textsc{SmartEmbed}~\cite{smartembed, gao2020checking} is the only tool that is specialized for calculating the similarity between smart contracts developed by Solidity at the time of writing this paper. According to their paper, \textsc{SmartEmbed} obtains excellent performance in calculating the similarity of Ethereum smart contracts and outperforms the traditional similarity-checking / clone-detection tools, e.g., Deckard~\cite{jiang2007deckard}. Thus, we use \textsc{SmartEmbed} to calculate the code similarity instead of using other similarity checking techniques, e.g., vanilla~\cite{xiong2015different}. 

\textsc{SmartEmbed} first converts a smart contract into a code embedding by parsing the AST of a smart contract (details see Section ~\ref{sec:related}) and then calculate the similarity between two contracts. The similarity metric is calculated as: $ Similarity (c_{1}, c_{2}) = 1 - \frac{Euclidean (e_{1}, e_{2})}{||e_{1}|| + ||e_{2}||}$, where $c_{1}$ and $c_{2}$ are two smart contracts; $e_{1}$ and $e_{2}$ are their corresponding code embeddings, which can be presented as  $e_{i} = \{w_{i_1}, w_{i_2}, ..., w_{i_n}\}$. Euclidean function is used to point the distance between $e_{1}$ and $e_{2}$, which is calculated by $Euclidean (e_{1}, e_{2}) = \sqrt{(w_{1_1} - w_{2_1})^{2} + (w_{1_2} - w_{2_2})^{2} + ... + (w_{1_n} - w_{2_n})^{2}}$.  Although the tool is aimed at finding bugs, their first step computes code similarity between the given smart contract and history contracts. We modified the source code of \textsc{SmartEmbed} to compute the similarity between two contracts. If their similarity is larger than 0.6,  they might be relevant and we assume the later created contract is the successor of the self-destructed contract. We also called this self-destructed contract as the predecessor contract of the successor contract.  We found 436 self-destructed contracts have their successor contracts with 1513 \textit{\textless predecessor contract, successor contract\textgreater} pairs. We note that 0.6 is a conservative threshold (original paper assumes similarity 0.95 are cloned); we might include many irrelevant pairs in our dataset, but it will not influence our result as we conduct a manual analysis in the subsequent step. Increasing the threshold can remove some irrelevant pairs to reduce the manual effort, but it might make us miss some true matching pairs. Besides, some token contracts might have duplicated code with high similarity, but it still will not affect the results. Because we will analyze the difference of the similar token contracts manually to identify whether the later created contracts are the successor contracts of the prior contracts. 

\subsubsection{Stage 3. Reason Generation:}
%\noindent{ \textbf{4.2.3 \, Stage 3: Reason Generation}}
 
 In stage 2, we found 436 contracts among 756 self-destructed contracts that have successor contracts. Note that the PS pairs we found in stage 2 might contain many false positives. For example, two contracts can obtain very high code similarity if they use many common open-source libraries, but they are not the valid PS pairs. Thus, we need manual analysis to remove these false positives. 

It is a time-consuming and error-prone process to analyze the predecessor contract and successor contract directly. To deal with these issues, we perform two steps. First, to reduce the manual effort, we use a tool named \textit{DiffChecker}~\cite{diffchecker} to help us find the difference between the predecessor and successor contract. The basic idea of \textit{DiffChecker}  is to use the Longest Common Substring (LCS)~\cite{LCS} algorithm to find the longest string (or strings) that is a substring (or are substrings) of two or more strings. We use \textit{DiffChecker} to highlight the difference to reduce manual efforts. Second, to increase the reliability of our results, we conduct open card sorting to summarize the reason why a smart contract self-destructed. Guided by previous works~\cite{chen2020defining, spencer2009card}, we create one card for each \textit{PS} pair. Each card highlights the difference between the two contracts.

\begin{figure}
	\begin{center}
		\includegraphics[width=0.95\textwidth]{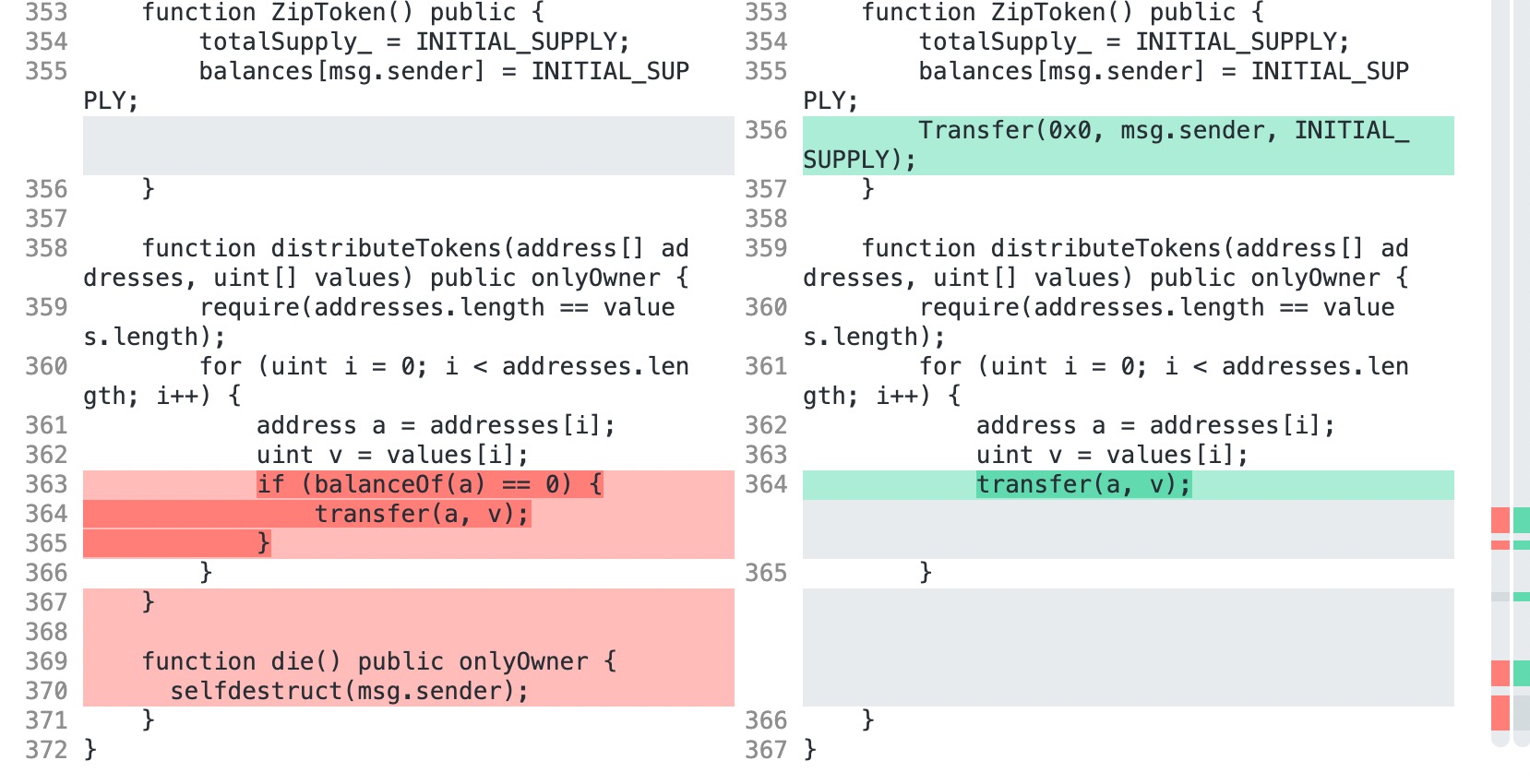} 
		\caption {An example considered in our card sort that contain a predecessor contract (0x3b96990a8ef293cdd37c8e1ad3d210a0166f40e1) and successor contract (0xedd7c94fd7b4971b916d15067bc454b9e1bad980).} 
		\label{Fig:diff}
	\end{center}
\end{figure}

The detailed steps of the open card sorting we used are: 

\quad \textbf{Iteration 1:} We randomly chose 20\% of the cards, and two developers with 3 years of smart contract development analyzed the difference of the code and discussed the reason why contracts self-destructed. They first quickly read the two contracts to identify whether they are relevant (The two contracts have similar functionalities). If they are irrelevant, the card will be discarded. Then, they carefully read the difference between the contracts and discuss the reason for this difference. For example, Figure~\ref{Fig:diff} is a real example of a predecessor contract (left) and its successor contract (right) in our dataset. The three differences between the two contracts are highlighted. First, the developer added a \textit{Transfer} event in Line 356 of the successor contract. Second, in the predecessor contract, Ethers can only be sent to an address whose balance is zero. This restriction was removed in the successor contract. Finally, the \textit{selfdestruct} function was also removed. According to our definition, all of these three modifications change the code representation of the contract. Thus, the reason for the Self-destruct is regarded as \textit{Functionality Changes}. For some contracts, it is not easy to find the reason for the self-destruct usage and they were omitted from our card list. All the reasons are generated during the sorting. 

\quad \textbf{Iteration 2:} The same two smart contract developers independently categorized the remaining 80\% of the cards into the initial classification scheme. Then, they compared their results and discussed disagreements. We used Cohen’s Kappa~\cite{kappa} to measure the agreement between the two developers. Their overall Kappa value is 0.84, indicating strong agreement.

We finally identified 5 reasons why contracts have been self-destructed. This information is shown in Table 1 and the detailed information is shown in the following subsection. 

\begin{table}
	\footnotesize
	\caption{Reasons of Self-destruct and their distributions among 340 self-destructed smart contracts.}
	\label{tab:reasons}
	\centering
	%\begin{tabular}{c | c}
	\begin{tabular}{p{85pt} | p{200pt} | p{40pt}}
		\hline
		Category & Description & Distribution\\
		\hline
		Functionality Changes & Adding, removing or changing Functionalities for upgrading contracts to respond new requirements. Functionality changes will change the code representation of a contract, e.g., Abstract Syntax Tree (AST), Control Flow Graph (CFG). & 156 (45.88\%) \\
		\hline
	%	Confusing Contract & Contract on Etherscan contains unused subcontracts, which might confuse users. Removing used parts to increase the readability. & 11 \\
	%	\hline
		Limits of Permission & Adding permission checks for the sensitive functions. & 25 (7.35\%)\\
		\hline
		Unsafe Contracts & Removing the security problems of the contracts. & 95 (27.94\%) \\
		\hline
		Unmatched ERC20 Token & Modifying the contract to make it follows the ERC 20 standard & 19 (5.59\%)\\
		\hline
		Setting Changes & Changing the variable or function states of the contracts, such as renaming a contract, changing the amount of ERC20 token supplement, changing a public function to private function. Setting Changes will not remove or add new code from a contract. & 56 (16.47\%) \\
	\end{tabular}	
\end{table}

\subsection{Reasons for Self-destruct} 

In this subsection, we give detail explanations of the 5 self-destruct reasons and their distribution in our dataset. 

%\noindent \textbf{4.3.1 Definitions.} 
\subsubsection{Definitions} 
The short descriptions of 5 self-destruct reasons are given in the first two columns of Table~\ref{tab:reasons}. Below we give a detailed description  each reason.	

\noindent \textbf{(1) Functionality Changes}:   Due to the immutability of smart contracts, it is not easy to upgrade smart contracts. However, during the entire smart contract life cycle, it is necessary for developers to add, remove or change some functionalities to respond to the new requirements. Functionality changes will change the code representation of a contract, e.g., Abstract Syntax Tree (AST), Control Flow Graph (CFG). According to our analysis, we find that functionality changes are the most common reason why smart contracts are self-destructed. When new requirements appear or some requirements are changed, some developers choose to deploy a new smart contract and the old version of the contract will be destructed.

\noindent \textbf{(2) Limits of Permission}: Ethereum is a permission-less network~\cite{chen2020maintaining} and anyone can call the functions of the contracts by sending a transaction. Thus, it is important to limit the access permissions for some sensitive functions. According to our analysis, we find some predecessor contracts do not check permissions of the callers in some sensitive functions, e.g., Ether transfer. Thus, everyone can execute the sensitive functions. In the successor contract, they add permission checks to limit access permissions.

\begin{figure}
	\begin{center}
		\includegraphics[width=0.95\textwidth]{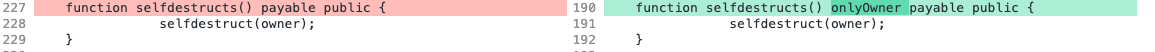} 
		\caption {A real diff example of Limits of permission in our dataset. Predecessor Address: 0xfa1d63b87f40c92d27bfb255419c1ea8c49086de; Successor Address: 0x64b09d1a4b01db659fc36b72de0361f2c6c521b1} 
		\label{Fig:diff-permission}
	\end{center}
\end{figure}

\textbf{Example}: Figure~\ref{Fig:diff-permission} is a real example of Limits of Permission. The predecessor contract does not limit the permission of calling the selfdestruct function. In the successor contract, the function adds a modifier \textit{onlyOwner} to check the permission. Thus, only the contract owner can call this function. 

% \begin{lstlisting}[caption={Example: Limits of permission},label=List:limits]
% //Predecessor Contract: 0xfa1d63b87f40c92d27bfb255419c1ea8c49086de 
% function selfdestructs() payable public {
% 	selfdestruct(owner);}
% //Successor Contract: 0x64b09d1a4b01db659fc36b72de0361f2c6c521b1 
% function selfdestructs() onlyOwner payable public {
% 	selfdestruct(owner);}
% \end{lstlisting} 

\noindent \textbf{(3) Unsafe Contracts}: Previous works~\cite{oyente,Zeus,Maian,Securify, chen2020defining} highlighted several security problems of smart contracts. For example, \textit{Oyente} highlighted four security issues, \textit{Zeus} described four security problems of smart contracts (see Section~\ref{sec:related}). We find many predecessor contracts contain security problems like \textit{reentrancy}, which can lead to Ether loss. Developers usually fix these security issues in the successor contracts.			

\begin{figure}
	\begin{center}
		\includegraphics[width=0.95\textwidth]{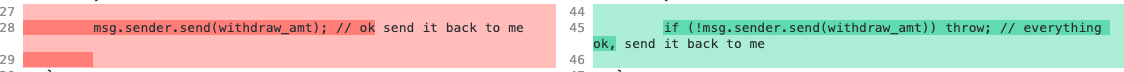} 
		\caption {A real diff example of Unsafe Contract in our dataset. Predecessor Address: 0x8b099bdcfea93faecfac13d0dbc1d08c4e1ec595; Successor Address: 0x17683235257f2089e3e4acc9497f25386a529507} 
		\label{Fig:diff-safe}
	\end{center}
\end{figure}

\textbf{Example}: Figure~\ref{Fig:diff-safe} is a real example of Unsafe Contracts. The predecessor contract does not check the return value of the  \textit{msg.sender.send()}, which might lead to security issues of the contract. In the successor contract, the function checks whether the Ether send is successful. If not, the transaction will be thrown. 

% \begin{lstlisting}[caption={Example: Unmatched ERC20 token},label=List:unmatched]
% //Predecessor Contract: 
% function transfer(address _to, uint _value); 
% //Successor Contract: 
% function transfer(address _to, uint _value) returns (bool success);
% \end{lstlisting} 

\begin{figure}
	\begin{center}
		\includegraphics[width=0.95\textwidth]{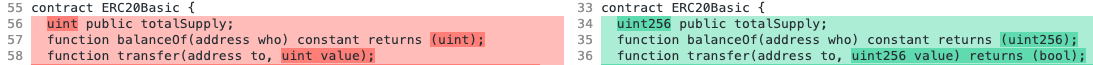} 
		\caption {A real diff example of Unmatched ERC20 token in our dataset. Predecessor Address: 0x848217a9569ca64fffba9d000cda05f9d2fa97f5; Successor Address: 0xf42230a7e21375c29648ae9544f7da394e20ead3 } 
		\label{Fig:diff-ERC20}
	\end{center}
\end{figure} 

\noindent \textbf{(4) Unmatched ERC20 token}:  ERC20~\cite{erc20}  is the most popular standard interface for tokens in Ethereum. If the implementation of token contracts does not follow the ERC20 standard strictly, the transfer between tokens may lead to errors. We find many predecessor contracts are token contracts but do not strictly follow the ERC20 standard, while their successor contracts do follow the standard.

\textbf{Example}: ERC20 requires a transfer function to return a boolean value to identify whether the transfer is successful. However, the transfer function in the predecessor contract in Figure~\ref{Fig:diff-ERC20} does not return anything. Users usually use third-party tools to manipulate their tokens and these tools capture token transfer behaviors by monitoring standard ERC20 method~\cite{chen2019tokenscope}. If the contract does not match the ERC20 standard, the token may fail to be transferred by third-party tools. In the successor contract, the return value of the \textit{transfer()} function is added.

\noindent \textbf{(5) Setting Changes}: Similar to \textit{Functionality Changes}, it is likely that developers will change some settings of smart contracts in response to new requirements. For example, the token-related contracts usually have some default values, e.g., total token supply, number of decimals, token name. Due to immutability, if developers want to change the total token supply, they have to destruct the contract and deploy a new one. The main difference between \textit{Setting Changes} and \textit{Functionality Changes} is that \textit{Setting Changes} will not add or remove code from contracts. Thus, the structure of AST and CFG should be the same between two contracts, if the reason for their selfdestruct is \textit{Setting Changes}.

\begin{figure}
	\begin{center}
		\includegraphics[width=0.95\textwidth]{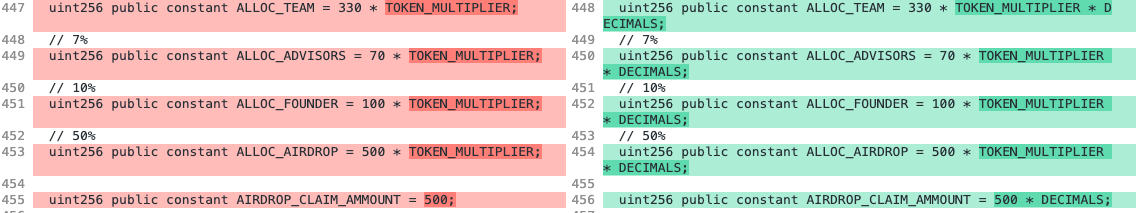} 
		\caption {A real diff example of Setting Changes in our dataset. Predecessor Address: 0xa41aa09607ca80ee60d2ce166d4c02a71860e5c5 ; Successor Address: 0x41c6af7b388e80030e63f2686dc2ff9bfd1267c9 } 
		\label{Fig:diff-setting}
	\end{center}
\end{figure} 

\textbf{Example}: Figure~\ref{Fig:diff-setting} is a real example of Setting Changes. The token supplement in the predecessor contract is too small. Thus, the successor contract changes the token supplement by multiple a value DECIMALS (DECIMALS equals to $10^{18}$ in the successor contract.)

%\begin{figure}
%	\begin{center}
%		\includegraphics[width=0.85\textwidth]{./figures/Fig5.pdf} 
%		\caption {Distribution of the reasons of self-destruct from 351 self-destruct smart contracts that can find their successor contracts.} 
%		\label{Fig:distribution}
%	\end{center}
%\end{figure} 

\subsubsection{Distribution}
%\label{sec:distribution}
%\noindent \textbf{4.3.2 Distribution.}
We use \textsc{SmartEmbed} to find the pair\textit{\textless predecessor contract, successor contract\textgreater}, and set the threshold value to 0.6, which might obtain some irrelevant pairs. After manually removing irrelevant pairs, we found 340 contracts (\textbf{some contracts have multiple self-destruct reasons. If multiple changes are found in the successor contract, we regard all of the changes as contributing to the self-destruction.}) for which we can identify the reason(s) why they are self-destructed and give the distribution of the five self-destruct reasons in the last column of table~\ref{tab:reasons}. 96 contracts cannot find the reason why they are self-destructed, and thus they are omitted from our dataset. 

It is clear that \textit{Functionality Changes, Unsafe Contract}, and \textit{Setting Changes} are the top three most popular reasons that lead to contracts destructed; the number are 156, 95 and 56, respectively. The number of the other two reasons are similar, there are 25 contracts destructed for \textit{Limits of Permission} and 19 for \textit{Unmatched ERC20 Token}.
%\textit{Confusing Contracts} is the least popular reason that leads to the destruct of contracts. However, we find that only 12 contracts are \textit{Confusing Contracts} in the predecessor contract, and 11 of them removed the redundant contracts in their successor version, which means 11 out of 12 developers realize their contract contains unused parts and it might confuse users. It shows this problem is also important. 

It should be noted that it is not easy to find all the security issues in our dataset. One the one hand, we only checked the security issues reported by \textit{Oyente, Zeus, Mythril, Securify, Maian}~\cite{oyente,Zeus,Maian,Securify, Mythril}. On the other hand, manually checking for security issues is very error-prone and time-consuming. To reduce the errors, we utilized tools by \textit{Oyente, Zeus, Mythril, Securify, Maian} and manually checked each contract it found. We first use the tools to check smart contracts. Then, two developers with 3 years of smart contract development experience manually identified whether the results are correct. If the reported results were different, they discussed to obtain the final result. Note that the code of \textit{Oyente, Mythril, Securify, Maian} can be found on Github, and we rerun the tool to get the result. We did not find the code of \textit{Zeus}, but \textit{Zeus} provides their evaluation results which inform whether a contract address contains vulnerabilities or not. Thus, we use their evaluation results directly. If the detected contract does not appear in their evaluation results, we regard Zeus as not finding any vulnerabilities in the contract.

	\section{RQ3: LifeScope: A Self-destruct Issues Detection Tool for Smart Contracts}
\label{sec:RQ3}

\subsection{Motivation}
In the previous section, we introduced five smart contract self-destructed reasons by comparing the difference between predecessor contracts and their successor contracts. Among these five reasons, \textit{Functionality Changes} and \textit{Change Setting} depend subjectively on the contract owner's requirements. Specifically, different developers might make different decisions of whether a smart contract should be self-destructed according to their requirements, even if the smart contracts are the same. It is thus hard to say these two reasons can affect the life cycle of smart contracts. However, smart contracts that contain the other three self-destruct reasons might have a short life span, as they can lead to unwanted behaviors of the smart contracts. Detecting whether a smart contract contains these self-destruct reasons might increase the life span of the contract. Manual analysis is time-consuming and error-prone. Therefore, designing a tool to detect whether a contract contains these self-destructed reasons before deploying them to the Ethereum is important. Security issues is a big concept. In the last section, we use the security vulnerabilities defined in previous works, e.g., \textit{Oyente, Zeus, Mythril, Security, Maian}~\cite{oyente,Zeus,Maian,Securify, Mythril}, to find unsafe contracts. These have already proposed several tools to detect security issues with high accuracy. The accuracy of \textit{Zeus} is almost 100\% according to their paper, and designing a more accurate and comprehensive security detecting tool is not the main target of this paper. In this case, we do not redevelop a tool to detect security issues introduced in these previous works.

\subsection{Approach}
\label{sec: lifescope_method}

We propose a tool named \textit{LifeScope} to detect the remaining two issues, i.e.,  \textit{Limits of Permission} and \textit{Unmatched ERC20 Standard} that can lead to contracts being destructed. Since the aim of \textit{LifeScope} is extending the life span of a smart contract by finding the self-destruct reasons, and smart contracts are immutable to be modified after deploying to the blockchain. Therefore, it is meaningless to detect the two self-destruct reasons through bytecode, although smart contracts are stored in the form of bytecode.

%\begin{figure*}
%	\begin{center}
%		\includegraphics[width=0.85\textwidth]{./figures/Fig2.pdf} 
%		\caption {Approach to Detect \textit{Unmatched ERC20 Token}} 
%		\label{Fig:erc}
%	\end{center}
%\end{figure*}

\begin{lstlisting}[caption={ERC20 Functions and Events},label=List:erc20]
function totalSupply() public returns (uint256)
function balanceOf(address _owner) public view returns (uint256 balance)
function transfer(address _to, uint256 _value) public returns (bool success)
function transferFrom(address _from, address _to, uint256 _value) public returns (bool success)
function approve(address _spender, uint256 _value) public returns (bool success)
function allowance(address _owner, address _spender) public view returns (uint256 remaining)
event Transfer(address _from, address _to, uint256 _value)
event Approval(address _owner, address _spender, uint256 _value)
\end{lstlisting} 

\begin{algorithm}
	\caption{Algorithm to Detect Unmatched ERC20 Token}
	\label{alg:erc20}
	\SetKwInput{KwOffInput}{Input}             
	\SetKwInput{KwOffOutput}{Output}      
	\SetAlgoLined
	% \KwResult{Write here the result }
	\KwOffInput{AST of a smart contract}
	\KwOffOutput{Is an Unmatched ERC20 Token}
	Extract $\langle fun_n, funInfo \rangle$ pair list from AST\;
	Extract $\langle event_n, eventInfo \rangle$ pair list from AST\;
	appearedFunc = 0\;
	legalFunc = 0\;
	legalEvent = 0\;
	isMachedERC20 = false\;
	\For{$fun_{i}, funInfo_{i} \in \langle fun, funInfo \rangle  \ pair \  list$}
	{
		\If{$fun_{i}$ is one of ERC20 Standard Function} {
			appearedFunc++\;
			\If{$funInfo_{i}$ is same to ERC20 Standard Function} {
				legalFunc++\;
			}
		}
	}
	\For{$event_{i}, eventInfo_{i} \in \langle event_, eventInfo \rangle \  pair \ list$}
{
	\If{$event_{i}$ is one of ERC20 Standard Event} {
		\If{$eventInfo_{i}$ is same to ERC20 Standard Event} {
			legalEvent++\;
		}
	}
}
\If{appearedFunc < 5}{
	return not\_ERC20\_Contract;
}

\If{legalFunc == 6 and legalEvent == 2}{
	isMachedERC20 = true;
}
return isMachedERC20;

\end{algorithm}

\textit{LifeScope} detects the self-destruct issues at source code level, which utilizes AST (abstract syntax tree) to parse the smart contracts and extract related information to detect \textit{Unmatched ERC20 Standard}. For \textit{Limits of Permission}, \textit{LifeScope} first transfers the contract to a TF-IDF representation and then utilizes machine learning algorithms to predict this problem.  These two problems are not only limited to contracts that contain the \textit{selfdestruct} function. Any smart contracts can be analyzed with \textit{LifeScope} to detect these two problems before deploying them to the Ethereum.

\subsubsection{Unmatched ERC20 token}
%\noindent \textbf{(1). Unmatched ERC20 Token}:
%Figure~\ref{Fig:erc} shows our approach to detect an \textit{Unmatched ERC20 token}; 
The method to detect the \textit{Unmatched ERC20 Token} is shown in Algorithm~\ref{alg:erc20}. The input is an AST (Abstract Syntax Tree) of a smart contract, which is generated by the \textit{Solidity} compiler~\cite{solc}. AST is a tree structure that contains the syntactic information of source code. By analyzing the AST, we can generate a list of pairs $\langle fun_n, funInfo \rangle$. $fun_n$ is the name of a function; $funInfo$ contains information about the function $fun_n$, i.e., parameter types and return types. ERC20 standard defines nine functions and two events. Among the nine functions, three are optional, and six are compulsory. The six compulsory functions and two events are shown in Listing~\ref{List:erc20}. We traverse all the functions in the smart contract. If the function name $fun_n$ is one of the six compulsory functions, we then compare whether the input parameter types and return types are the same with the ERC20 standard. For example, if a contract contains a function named \textit{transfer}, we then check whether this function contains two parameters and their types are \textit{address} and \textit{uint256}, respectively. Besides, the function should have a return value, and the type of the return value is \textit{bool}. We use the same method to check  contract events. Finally, if all the six compulsory functions and two events appear in the contract, this is a matched ERC20 smart contract. We follow the previous work~\cite{frowis2019detecting}, if less than five ERC20 functions appear in a smart contract, we regard it as not an ERC20 smart contract. Otherwise, it is an unmatched ERC20 token. Note that Solidity allows abstract functions or interfaces which do not have code implementation. We only consider a function as an ERC20 function if it has complete implementation. Besides, the bytecode of a public storage variable is the same as a view function without any input parameters and only has one return value. Thus, the function \textit{totalSupply()} can also be represented as a global variable, i.e., \textit{uint totalSupply = amount;} If a smart contract does not contain a function named \textit{totalSupply()}, but contains a public variable named \textit{totalSupply}, it will also be regarded as a matched ERC20 function.

\begin{figure*}
	\begin{center}
		\includegraphics[width=0.85\textwidth]{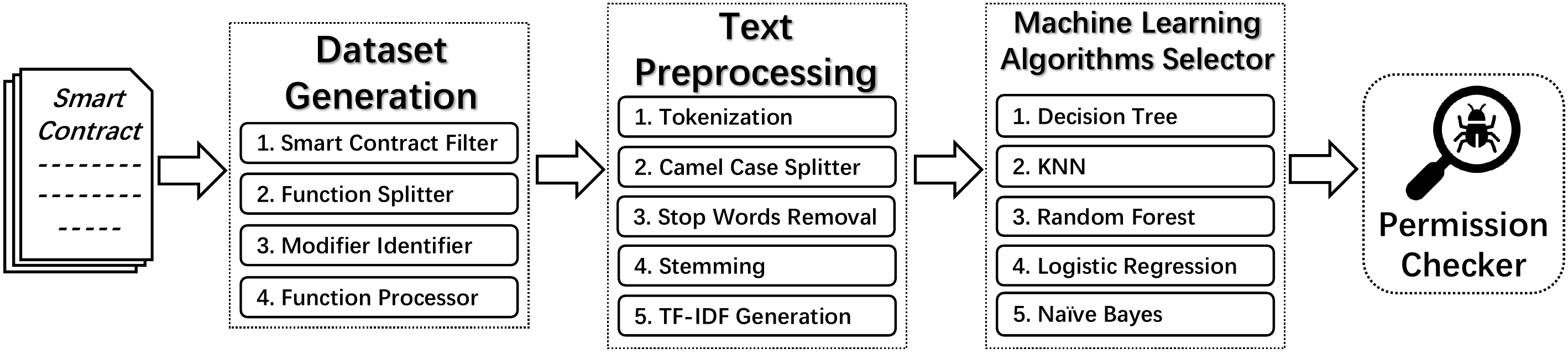} 
		\caption {Approach to Check Permissions} 
		\label{Fig:permissions}
	\end{center}
\end{figure*}

\subsubsection{Limits of Permission}
\label{sec:permission}
%\noindent \textbf{(3). Limits of Permission}: 
It is hard to prescribe functions need to check their permissions. Therefore, It is not easy to detect this issue by using programming analysis methods. We utilize a machine learning method to predict whether a function needs to check for its caller permission. Figure~\ref{Fig:permissions} describes the overall architecture that we used. The method contains three parts, i.e., \textit{Dataset Generation}, \textit{Text Preprocessing} and \textit{Machine Learning Algorithm Selector}.

\textit{\textbf{(a). Dataset Generation:}} The aim of this step is to extract pairs\textit{<func, permission>} from smart contracts. In each pair, \textit{func} is the source code of a function in the smart contract, and \textit{permission} means whether the function needs to check the caller's permission. Since security vulnerabilities are ubiquitous in smart contracts on Ethereum~\cite{oyente, chen2018infocom, Zeus, Maian},  some alive contracts might also miss checking the permissions of the functions. This situation gets even worse in self-destructed contracts, as the reason for the destruct might be missing permissions. Therefore, it is not reliable to use these contracts as our ground truth. To ensure the correctness of our dataset, we should use contracts that correctly check their permissions for the contracts.  However, it is not easy to ensure the correctness of the contract, and manually check whether a function needs to check its permission is also error-prone and subjective. 

To obtain the dataset, we first rank all the alive verified contracts by their transaction numbers. We then choose all of the contracts whose transaction numbers are larger than 500, as transactions can be regarded as test cases for the contract. Previous works check the transaction input to detect malicious attacks~\cite{ferreira2019aegis, wang2019vultron, chen2020soda}. The more the number of normal running transactions a contract has, the less likely the contract has permission problems.  After this step, 5,986 contracts remained. Financial gain is an important motivation behind the attacks on Ethereum smart contracts~\cite{chen2020maintaining}. We finally find 1 Ether can get an appropriate number of the dataset (875 smart contracts with 29,313 functions). Thus, we choose contracts whose balance have more than 1 Ether (1 Ether worths about \$1400 at Feb. 2021). The sensitive information of smart contracts, e.g., contract bytecode, balance, are visible to the public. Finally,  875 contracts whose balance has larger than 1 Ether and transaction numbers larger than 500 transactions remain. %As reported by previous works, attacks are prevalent on the Ethereum platform~\cite{zou2019smart, chen2020defining}. A permission-less contract with high balance can easily be the target of attackers.

%In this case, we make a hypothesis that if a smart contract has permission problems and the problems are found by the contract owner or attackers, the Ethers on the contract will be transferred. Therefore, the balances on the contract will return to zero. The more the Ethers on the contracts, the higher probability that the permission problems be found by contract owners or attackers. In other words, the more Ethers on the contracts, the lower risk of having permission problems in the contracts. 

After getting these contracts, we first remove the comments on the contracts and then split the contracts into functions. We obtain 29,313 functions from these 875 smart contracts. We then need to identify whether these functions contain modifiers, as the permission is usually checked by the modifier. If a function contains modifier, we remove the modifier from the function. For example, the original function is \textit{function transferMoney(address addr) onlyOwner\{\}}. We remove the modifier \textit{onlyOwner} from the function and only use \textit{function transferMoney(address addr)\{\}} to training the machine learning model. In our dataset, we obtain 29,313 functions, with 4,393 of them needing to check permissions.  

\textit{\textbf{(b). Text Preprocessing:}} Before training the machine learning module, we need to transfer each processed function into a set of bag-of-words (BoW). The dataset is split into a training set and a test set. Both of them are first processed by the following steps: \textbf{(i) Tokenization}: Each processed function is divided into a list of words by punctuation and space that usually do not contain any information. For example, \textit{function transferMoney(address addr)\{\}} will be transferred into "function", "transferMoney", "address" and "addr" \textbf{(ii) Camel Case Splitter:} We separate function names, variable names and identifiers according the rules of Camel Case~\cite{CamelCase}. For example, "transferMoney" will be separated into "transfer" and "Money". Then, we transfer all the words to their lower case.\textbf{ (iii) Stop Words Removal:} Stop words means meaningless words (e.g., "to", "as", "is"). We adopt NLTK (a python library) stop words list in this step. We also remove tokens of less than 3 characters.  \textbf{ (iv) Stemming:} this step is used to transfer words into their stem form. For example, "running" is replaced by "run". In this paper, we use Porter's stemmer~\cite{porterStemmer} to transfer the words. 

After these four steps, we use TF-IDF (term frequency - inverse document frequency)~\cite{tfidf} to represent each processed word in the function. (Noticed that to make the result more reliable, the BoW model is only built by the training set.) This is described as: %$w_{i,j} = tf_{i,j} * log(\frac{\# of functions}{df_{i}} )$. 

\begin{eqnarray}
w_{i,j} = tf_{i,j} * log(\frac{\# of functions}{df_{i}} )
\end{eqnarray}

Here, $w_{i,j}$ is the weight of the word \textit{i} in the function \textit{j}. $tf_{i,j}$ is the term frequency of word \textit{i} in the function \textit{j}. $df_{i}$ is the number of functions that contain word \textit{i}. Finally, each function is represented as $fun_j = (w_{1,j}, ... w_{i,j}, ..., w_{n,j})$. 

\textit{\textbf{(c). Machine Learning Algorithms Selection:}} Checking the permission of functions is a binary classification problem. We tried five popular machine learning algorithms to find an appropriate algorithm for predicting the permission and use the algorithm that obtains the best F-Score for this task. The five algorithms are \textit{Decision Tree, KNN, Random Forest, Logistic Regression}, and \textit{Naive Bayes}. We use a python library named sklearn~\cite{sklearn} with its default configuration to implement these five algorithms.

\subsection{Evaluation For Unmatched ERC20 Token}
%\label{sec:res_tool}

\subsubsection{Dataset}
The dataset for \textit{Unmatched ERC20 token} consists of two parts, i.e., smart contracts with and without \textit{selfdestruct} function. The dataset with \textit{selfdestruct} function has 756 self-destructed contracts, which is the same dataset introduced in RQ2. Among these contracts, 127 of them are discarded due to the unsupported compiler version. (LifeScope supports Solidity compiler versions that are equal to or higher than 0.4.25). So, there are finally 629 self-destructed contracts in our dataset. The two researchers manually labeled the dataset and found 164 of them are ERC 20 token contracts. In these 164 ERC20 token contracts, 70 (11.13\%) of them are \textit{Unmatched ERC20 tokens}.

For the dataset without \textit{selfdestruct} function, we use an open source dataset proposed by our previous work~\cite{chen2020defining}. The dataset contains 587 smart contracts, which has the ground truth of \textit{Unmatched ERC20 tokens}. The ground truth is labeled by two experienced researchers. They first analyzed the contracts independently, and then discussed any differences, which ensured the correctness of the dataset. Since the dataset is randomly selected, some contracts contain the \textit{selfdestruct} function and a low compiler version. After removing the incompatible contracts, 358 contracts remained. Among these 358 smart contracts, 141 are ERC20 token contracts, and 36 (25.53\%) of them have an Unmatched ERC20 token.

\subsubsection{Result} For the dataset with \textit{selfdestruct} function, \textit{LifeScope} finds 70 \textit{Unmatched ERC20 token}, with 0 false positive and negative. For the dataset without \textit{selfdestruct} function, LifeScope finds all 36 Unmatched ERC20 token, with 0 false positives and negatives. The results show that \textsc{LifeScope} can also detect the  \textit{Unmatched ERC20 token} problem in both smart contracts with and without \textit{selfdestruct} functions.

\subsubsection {Comparison:} \textit{TokenScope}~\cite{chen2019tokenscope} is a novel transaction based tool to identify the inconsistency of ERC20 tokens. It can identify whether a contract is a legal ERC20 token by investigating its transactions. In this paper, we re-execute \textit{TokenScope} and use the same dataset to compare with LifeScope.	

The dataset with a \textit{selfdestruct} function contains 164 ERC20 contracts. Among these contracts, 94 are legal ERC20 contracts, and 70 have an unmatched ERC20 Token. \textit{TokenScope} correctly predicts 88 contracts are legal ERC20 tokens, and 49 are unmatched ERC20 tokens. However, it mistakenly predicts 6 contracts are not legal ERC20 tokens, and 21 are ERC20 tokens.

The dataset without \textit{selfdestruct} function contains 141 ERC20 contracts. Among these contracts, 105 are legal ERC20 contracts, and 36 are unmatched ERC20 Tokens. \textit{TokenScope} correctly predicts 96 contracts are legal ERC20 tokens, and 17 are unmatched ERC20 tokens. However, it mistakenly predicts 9 contracts are not legal ERC20 tokens, and 19 are ERC20 tokens.

\textit{TokenScope} needs transactions to identify whether a contract has legal ERC20 tokens. However, since some contracts only have a signal transaction, this leads to the false positives of \textit{TokenScope} (mistakenly predicting a contract is not an ERC20 token). Besides, we find \textit{TokenScope} cannot check the return value of a function. Specifically, ERC20 standard requires the transfer function to return a boolean value, while \textit{TokenScope} cannot identify whether a function has a return value, which leads to false negatives (mistakenly predicting a contract is an ERC20 token). 

In conclusion, \textit{LifeScope} performs better than \textit{TokenScope}. 

Etherscan also can identify whether a contract is an ERC20 contract. We did not compare the result with Etherscan because we have different standards to define whether a contract is an ERC20 token. ERC20 standard defines nine functions and two events. Among the nine functions, three are optional, and six are compulsory. For our paper, we use a definition in previous work~\cite{frowis2019detecting}: if less than five compulsory ERC20 functions appear in a smart contract, we regard it as not an ERC20 smart contract. A smart contract is defined as an unmatched ERC20 contracts only if it has more than or equal to five compulsory ERC20 functions, and some of these functions do not follow ERC20 standards. However, Etherscan has a different definition for ERC20 contracts. Etherscan regards a contract that has ERC20 related transactions as an ERC20 contract even though it only contains one ERC20 function, e.g., \textit{transfer()} and has related transactions. Thus, the result of Etherscan will be very different from our method, and it is the reason why we did not choose it as our comparison method.

\subsection{Evaluation For Limits of Permission}
%\label{sec:res_tool}

\subsubsection{Dataset}

The method that was used to generate the dataset of \textit{Limits of Permission} is introduced in section~\ref{sec: lifescope_method}. To better evaluate the results, we use a cross-validation method of training-testing sets. First, we divided our dataset into 10 parts of equal sizes. Then, we conduct the training using 7 parts of the dataset and 3 parts for testing. Specifically, we give an ID (0 to 9) to each part. In the first round, the parts with ID 0 to 2 are the testing sets, and 3-9 are the training sets. In the last round, the parts with ID 9, 0, 1 are the testing parts, and the remaining parts are the training set. We continue this process 10 times. Finally, we report the average results. There are around 20,328 case in our training sets and 8985 cases in our testing test.

\subsubsection{Evaluation Methods and Metrics}
We use five measurements to evaluate the results, i.e., precision, recall, F1-Measure, accuracy, and AUC. Precision, Recall, F-measure, and Accuracy can be calculated as: $\frac{\#TP}{\#TP + \#FP}$, $\frac{\#TP}{\#TP + \#FN}$, $\frac{2 \times P \times R}{ P + R}$, $\frac{\#TP + \#TN}{\#TP + \#TN + \#FN + \#FP} $, respectively. TP (true positive) indicates the number which correctly predicts a function needs to add a permission check. TN (true negative) indicates the number which correctly predicts a function does not need to add a permission check. FP (false positive) and FN (false negative) indicate the number which incorrectly predicts that a function needs or does not need to add a permission check. AUC (area under the curve) is calculated by plotting the ROC curve (receiver operator characteristic). 

\begin{table}
	%\small
	\footnotesize
	\caption{Results of Predicting \textit{Limits of Permission} by using Five Machine Learning Algorithms} 
	\label{tab:LoP_Algorithems}
	\centering
	\begin{tabular}{c | c | c | c | c | c }
		\hline
		& Precision & Recall& F-Measure & Accuracy & AUC\\
		\hline
		Decision Tree & 78.91\% & 77.09\% & 77.89\%  & 93.45\%  & 0.8673 \\
		\hline
		KNN & 72.75\% & 50.40\% & 59.50\% & 89.71\& & 0.7352\\
		\hline
		Random Forest& 83.73\% & 70.82\% & 76.05\% & 94.88\% & 0.8499 \\
		\hline
		Logistic Regression&84.11\% & 52.78\% & 64.82\% & 91.40\% & 0.7551 \\
		\hline
		Naive Bayes &  23.33\% & 88.43\% & 35.66\% & 52.15\% & 0.6708\\
		
	\end{tabular}  
\end{table}

\begin{table}
	%\small
	\footnotesize
	\caption{Results of Predicting \textit{Limits of Permission} for 10-fold cross-validation by using Decision Tree} 
	\label{tab:LoP_Res}
	\centering
	\begin{tabular}{c | c | c | c | c | c | c | c | c | c | c | c}
		\hline
		 & R1& R2& R3& R4& R5& R6& R7& R8& R9& R10 & AVG \\
		\hline
		Precision & 79.92\% & 79.08\% & 77.45\% & 76.76\% & 77.14\% & 77.04\% & 79.20\% & 77.32\% & 82.96\% & 82.19\% & 78.91\%\\
		\hline
		Recall & 69.08\% & 73.14\% & 77.28\% & 88.19\% &82.50\% &78.40\% &81.34\% &76.93\% &76.84\% &73.39\% &77.09\%\\
		\hline
		F-Measure& 74.10\% &76.00\% &77.37\% &79.29\% &79.73\% &77.72\% &80.26\% &77.13\% &79.78\% &77.54\% &77.89\% \\
		\hline
		Accuracy& 92.35\% &93.16\% &93.72\% &93.97\% &94.01\% &93.45\% &93.88\% &93.33\% &93.72\% &92.95\% &93.45\% \\
		\hline
		AUC & 0.8291 & 0.8489 & 0.8682 & 0.8896 & 0.8921 & 0.8721 & 0.8874 & 0.8653 & 0.8690 & 0.8511 & 0.8673\\

	\end{tabular}  
\end{table}

\subsubsection{Result} Table~\ref{tab:LoP_Algorithems} shows the result of predicting \textit{Limits of Permission} by using five machine learning algorithms.  We found that the \textit{Decision Tree} algorithm obtains the best F-Score for this task. Therefore, we finally use the \textit{Decision Tree} to predict whether a function needs to check for its callers' permission.  

The detailed results of predicting \textit{Limits of Permission} for 10-fold cross-validation by using Decision Tree are shown in Table~\ref{tab:LoP_Res}. Our method obtains 78.91\% of precision, 77.09\% of recall, 77.89\% of F-measure, 93.45\% of Accuracy, and 0.8673 of AUC. Prior works~\cite{fan2018chaff, fan2019impact} suggest that a classifier performs reasonably well if its AUC is larger than 0.7.

\subsubsection{Comparison} To evaluate the performance of \textit{LifeScope} in detecting \textit{Limits of Permission}, we design a keywords-based method to identify whether a function needs to check its permission. Following we describe the details of the comparison method. 

\noindent \textbf{Feature Selector:} We use decision tree to predict whether a function needs to add a permission check, which enable us to know 
which words lead to a function being classified as needing permission check. Guided by previous works~\cite{huang2018identifying, sebastiani2002machine, yang1997comparative}, we employ a widely used feature selection technique named Information Gain, to select useful features in our prediction.

Our dataset can be denoted as $F = {(F_1, L_1), (F_2, L_2), ..., (F_n, L_n)}$, where $F_i$ represents the $i^{th}$ function, and $L^i$ is the label, which means whether the $F_i$ needs to check its permission ($t$) or not ($\overline{t}$). The word vector of $F_i$ is represented as $F_i = \{w_1, w_2, ..., w_n\}$, where $n$ represents the number of different words appeared in $F_i$, and $w_i$ represents the $i^{th}$ words. There are four relationships between the word $w$ and a function $F_i$. 

1. $(w, t)$: function $F_i$ contains the word $w$, and the function needs to check its permission.
 
2. $(w, \overline{t})$: function $F_i$ contains the word $w$, but the function does not need to check its permission.
 
3. $(\overline{w}, t)$: function $F_i$ does not contain the word $w$, but the function needs to check its permission.
 
4. $(\overline{w}, \overline{t})$: function  $F_i$ does not contain the word $w$, and the function does not need to check its permission. 

Based on these relationships, the information gain (IG) of word $w'$ and label $t'$ is defined as: 

\begin{eqnarray}
IG(w, t) = \sum\limits_{t' \in \{t, \overline{t}\}}\sum\limits_{w' \in \{w, \overline{w}\}}p(w', t') \times log\frac{p(w', t')}{p(w') \times p(t')}
\end{eqnarray}

In the equation, $p(w', t')$ is the probability of the word $w'$ in a function with label $t'$. $p(w')$ means the probability of word $w'$ in a function and $p(t')$ represents the probability of a function with label $t'$.  

\noindent { \textbf{Behaviors of permission check:}} Information gain reflects the amount of information required for predicting a label (needs or does not need to check the permission). The higher IG score a word has, the more important the word to distinguish the label. We rank the IG score of each word and list the top 50 words with highest information gain score in Table~\ref{tab:ig}. The first line has the highest IG score, and the score decreases from left to right in each line. Note that a word with a high IG score means it has a high contribution to predicting the label, not necessarily mean that it indicates whether a function needs permission check. Thus, we manually check the word in Table~\ref{tab:ig} to find useful behaviors. 

From the words, we then summarize four kinds of behaviors that needs to check functions' permission: 

(1). \textbf{Ether Transfer.} Specifically, the words ``msg", ``sender", ``transfer", and ``eth" are related to Ether transfer methods on Ethereum, i.e., \textit{msg.sender.transfer(eth)}; the word ``withdraw" reflects a behavior related to withdraw balance. 

(2). \textbf{Sensitive states change}, e.g., changing permission owners, increasing or decreasing total supply of tokens, stopping or starting the functionalities. Specifically, the words ``ownership", ``administr", ``mint" and ``renounce" are all related to sensitive states of a contract.

(3). \textbf{Inline assemble and selfdestruct function.} Ethereum provides some functionalities which need to limit permission. Specifically, the words ``mload" and ``assemble" are related to the inline assemble. The word ``selfdestruct" is related to the \textit{selfdestruct} function.

(4).\textbf{ Emergency Management.} Smart contracts are difficult to be modified once deployed. Thus, there are some functions used to handle the emergency situations. Specifically, the words ``emerge", ``pause", ``stop", "unpause", ``selfdestruct" are related to emergency management.

We use the top 50 words with the highest IG score to find four behaviors that need to check functions' permission. To prove the completeness of the four behaviors, we also manually check the top 51-100 words with the highest IG score. We find there are no additional behaviors that can be found. All the words can be classified into these four categories.

\begin{table}
	%\small
	\footnotesize
	\caption{Top 50 words with highest information gain score.} 
	\label{tab:ig}
	\centering
	\begin{tabular}{c | c | c | c | c | c | c  | c | c | c}
		\hline
		ownership & transfer & name & oracle & pidx & gen & address & pause & human & round \\
		\hline
		stop & unpause & assert & online & data & addr & core & log & compress & assemble \\
		\hline
		withdraw & gap & event & mint & mask & eth & within & msg & transaction & mload \\
		\hline
		earn & emerge & spender & code & sender & renounce & determine & calc & sqrt & team \\
		\hline
		propose & math &  divest & div & selfdestruct & filter & administr & confirm & investor & sale \\
		
	\end{tabular}  
\end{table}

\noindent { \textbf{Result of the Comparison Method:}} In the last part, we summarize four kinds of behaviors that need to check functions' permission according to the top 50 words with the highest IG sore. According to these words, we design a simple keywords based method to identify whether a function needs to check its permission. Specifically, if a function contains one of the key word ``msg.sender.transfer", ``ownership", ``administr", ``mint",  ``renounce", ``mload", ``assemble" ``emerge", ``pause", ``stop" and "unpause", we assume the function needs to check its permission. Notice that we merge some keywords, e.g, ``msg", ``sender", ``transfer" to ``msg.sender.transfer" as ``msg.sender" are widely used in Ethereum. We finally obtain 31.53\%  precision, 11.41\% recall, 16.75\% F1-measure, and 83.01\% accuracy, which are much worse as compared to our machine learning based method (78.91\% precision, 77.09\% recall, 77.89\% F1-Measure, and 93.45\% accuracy).

	\section{Discussion}
\label{lab:discussion}
We first summarize the key implications of our work for researchers, practitioners, and educators. Then, we give 6 suggestions on how to better use \textit{selfdestruct} function according to the feedback of the survey. Finally, we summarize the main threats of validity.

\subsection{Implications}
\subsubsection{For Researchers}
%\noindent \textbf{6.1.1 For Researchers:}  
\textbf{\textit{Research Guidance. }} In this paper, we found 5 reasons why smart contracts destructed by comparing the difference between self-destructed contracts and their successor contracts. With the increasing number of smart contracts, researchers can apply our methods to find more problems that can affect the life span of smart contracts. Our study focuses on Ethereum smart contracts, but many other blockchain platforms also support the running of smart contracts, e.g., Ethereum Classic~\cite{ETC}, Expanse~\cite{Expanse}. Both Ethereum Classic and Expanse were created by the hard fork of Ethereum, but currently they are independent blockchain systems with many years of development. Both of them support the running of smart contracts based on EVM and support the \textit{selfdestruct} function. There might be some different reasons for the self-destruct; researchers can use our methods to identify the reasons on these platforms.

%We also designed a new tool, \textsc{LifeScope}, to detect three self-destruct reasons from the source code of smart contracts. However, the source code of smart contracts is not always visible to the public, while the bytecode is always free to check in Ethereum. Therefore, researchers can pay more attention to detect the problems at a bytecode level. Using \textit{Limits of Permission} as an example, \textsc{LifeScope} uses machine learning to predict whether a function need to check its permission. However, \textsc{LifeScope} needs users to provide it source code, which limits its usage. In Ethereum, smart contracts usually call other contracts, while the callee contract might not open up their source code. In this case, if researchers can check whether the functions on callee contracts has correct permission check, some financial loss might be avoided.

%\textbf{ \textit{Behavior vs. Perception~\cite{devanbu2016belief}.}} 

\subsubsection{For Practitioners}
%\noindent \textbf{6.1.2 For Practitioners:} 
Our work is the first that uses an online survey to collect feedback from smart contract developers on why they include or exclude \textit{selfdestruct} function. Their feedback shows that adding a \textit{selfdestruct} function can help developers transfer Ethers when emergency situations happen. However, using this function can also lead to several problems. To address the drawbacks of adding a \textit{selfdestruct} function, we give 6 suggestions in the next section. These  can help developers better use the \textit{selfdestruct} function in their contracts. Smart contract developers can develop a smart contract according to our suggestions and open source the code for other developers to use. We also summarized 5 common reasons why contracts self-destructed and developed the \textsc{LifeScope} tool to detect 2 self-destruct reasons. Removing these problems might extend the life span of smart contracts.   

\subsubsection{For Educators}
%\noindent \textbf{6.1.3 For Educators:} 
\textit{selfdestruct} is an important feature of smart contracts. However, most blockchain tutorials focus on teaching how to develop smart contracts and knowledge about blockchain~\cite{chen2020defining}. Educators should pay more attention to these unique functions of smart contacts. For example, educators should mention the importance and drawbacks of adding a \textit{selfdestruct} function when they introduce this function. The feedbacks of our survey in RQ1 can provide good materials for them.

\subsection{Towards More Secure selfdestruct Functions}
\label{sec:suggestion}
In Section~\ref{sec:RQ1}, we summarized six reasons why smart contract developers exclude the \textit{selfdestruct} function from their contracts. In this part, we give six suggestions about how to better use \textit{Selfdesturct} function according to the summarized worries.

\noindent \textbf{Suggestion 1. Limit Usage Scenario:} Adding \textit{selfdestruct} function can increase the complexity of the development and risk of attacks.  Some smart contract developers claim that the \textit{selfdestruct} function is mainly used to remove the code and transfer Ethers. However, they do not need this function if there is no Ether in their contracts. Removing the contracts from the blockchain is also unattractive to them. Even if their contracts are attacked and controlled by attackers, they can discard the old contracts and deploy a new version.  To reduce the risk and the workload of development, a \textit{selfdestruct} function is better to add in contracts that contain Ethers.

\noindent \textbf{Suggestion 2. Permission Check:} Calling a \textit{selfdestruct} function can lead to irreversible consequences. The contract has to check the permission of the caller in each transaction. A common method to check the permission is recording the owner's address in the constructor function. Then, checking whether the caller is the owner in each transaction.

\noindent \textbf{Suggestion 3. Distribute the Rights and Modularization: } The trust concern is an important reason why developers exclude a \textit{selfdestruct} function. The trust concern contains two parts according to the feedback, i.e., human related and code related concern.

For human related concern, users might worry that the owner of the contract can destruct the contract and transfer all the balance if the contract has a \textit{selfdestruct} function. For example, the \textit{gambling} contract shown in Listing 1 claims that users can transfer 1 Ether to the contract. When the contract receives 10 Ethers, the contract will choose one user as the winner and transfer 9 Ethers to the winner as a bonus. However, if the contract has a \textit{selfdestruct} function, users might worry that the owner might transfer the money out at any time.

To reduce this kind of concern, the owner could build a DAO (Decentralized autonomous organization) for the contracts. In a DAO system, the DAO controls the contracts, and the users of the contracts control the DAO by using digital tokens which give them voting rights. The users who hold tokens (voting rights) can submit a proposal, e.g., executing the \textit{selfdestruct} function. Then, the proposal will be checked by a group of volunteers called ``curators" to check the legality of the submitted proposal and the identity of the submitter. Finally, the users who owned the DAO tokens vote to accept or reject the proposal.

%To reduce this kind of concern, we suggest the owner should distribute the rights of calling the \textit{selfdestruct} function. For example, all the users who transfer the Ethers to the contract should have the right to vote whether the contract should be killed. The voting steps can follow some consensus protocols, e.g., PoS~\cite{PoS}, DPos~\cite{DPoS}. Using PoS as an example, the more Ethers people contribute to the contracts, the more votes they have. This can increase the cost of attackers who want to stop the execution of a \textit{selfdestruct} function.

However, adding a DAO pattern to the contract can increase code complexity, which is also a big concern according to our developer survey feedback. With the increase of code complexity, the probability of containing security vulnerabilities is also increasing, which leads to code related trust concerns for developers. To address these two concerns, we suggest that smart contract developers can open source and modularize this part of their code (DAO patterns) to a library. Other developers can then help polish the code together, and can make the code easier to use in the future.

\noindent \textbf{Suggestion 4. Delay Self-destruct Action: } The Ethers sent to a self-destructed contract will be locked forever, which increases the risk of using \textit{selfdestruct} function. As we described in Section~\ref{sec:RQ2-Motivation}, the contract owner might find it difficult to inform all the users in a short time after the contract self-destructed. In this case, some users might send Ethers to the self-destructed contract and this may lead to financial loss. To address this problem, we suggest that the contract can delay the self-destruct action and throw an event to inform the users that the contract will self-destructed in the near future. On the one hand, delaying self-destruct action can give time for voting (Suggestion 3). On the other hand, it can provide time to inform users that the contract will be destructed.

\noindent \textbf{Suggestion 5. Pause Functionality: } The options in Suggestion 3 and 4 require time to implement. However, when a contract is being attacked, any delay might lead to enormous financial loss. In this case, pausing the functionality when performing the methods described in Suggestion 3 and 4 are important. OpenZeppelin provides a \textit{Pausable} contract template~\cite{Pausable-Code, Pausable-Doc}, which can be easily used through inheritance. The three functions and two modifiers are shown in Listing~\ref{List:pause}. The two modifiers, i.e., \textit{whenNotPaused} and \textit{whenPaused}, can be added to control its states. Specifically, \textit{whenNotPaused} makes functions callable only when the contract is not paused, and \textit{whenPaused} makes a function callable only when the contract is paused. The state of the contract is obtained by a boolean value named \textit{\_paused}, which can change its state by the function \textit{\_pause()} and \textit{\_unpause()}.

\begin{lstlisting}[caption={OpenZeppelin Pausable contract},label=List:pause]
modifier whenNotPaused() { require(!paused(), "Pausable: paused"); _;}
modifier whenPaused() { require(paused(), "Pausable: not paused"); _;}
function paused() public view virtual returns (bool) { return _paused;}
function _pause() internal virtual whenNotPaused { _paused = true;}
function _unpause() internal virtual whenPaused { _paused = false;}
\end{lstlisting}

\noindent \textbf{Suggestion 6. Refund Values: } 
Some smart contracts might store the values of users. For example, users might hold tokens in ERC20 contracts. Thus, destructing the smart contracts will lead to financial loss of users. Before executing the \textit{selfdestruct} function, the contract owners should refund users' assets.

\subsection{Tokens of Destructed Unmatched ERC20 Contracts}
Unlike other kinds of contracts, ERC20 contracts usually store values (tokens) of users. Destructing the contract and transferring the balance is not enough for ERC20 contracts, as the tokens will be locked with the execution of the \textit{selfdestruct}  function. We investigate the transaction of 70 destructed unmatched ERC20 tokens introduced in Section 5.3 and find that 53 of them contain transactions that call \textit{transfer()}, which means there are token transfers in these contracts. However, only 3 contracts have refund-related transactions. These three contracts pay back a certain amount of Ethers according to the tokens that users have. For the other 50 smart contracts, the contract owners destruct the contracts directly without considering the benefits of users. This situation might suggest that ERC20 tokens with \textit{selfdestruct}  function might have a high risk for users.

\subsection{Inconsistency}
\subsubsection{\textit{selfdestruct} functions on Etherscan vs. Survey Results.}
 In this paper, we collected 54, 739 open-source smart contracts but only found 5.1\% of them contain the \textit{selfdestruct} function. However, in our survey, 38.82\% of the developers claim that they will add \textit{selfdestruct} functions to their contracts. There is an inconsistency between the practitioner's perception and their behavior (38.82\% vs. 5.1\%). The inconsistency might indicate that many developers admit the importance of \textit{selfdestruct} function but they give up on adding it during the actual development process. The reasons why developers do not add the \textit{selfdestruct} functions in actual developing process might have already been included in Section 3.3.2. Therefore, designing guidelines for using the \textit{selfdestruct} function might be helpful. Future work can aim to design guidelines, development models or tools to address these problems. We also give five suggestions in the Section~\ref{sec:suggestion}.

%In Section 4.3.2, we use 351 self-destruct smart contracts to give the distribution of the six self-destruct reasons. For 351 smart contracts we can find their successor contracts, which means developers destroyed the old contracts and deployed a new contract. There are 11 and 19 contracts in these 351 contracts that contain defects of \textit{Confusing contracts} and \textit{Unmatched ERC20 Tokens}, respectively. In Section~\ref{sec:res_tool}, we found that the number of \textit{Confusing contracts} and \textit{Unmatched ERC20 Tokens} in all 657 self-destruct contracts are 12 and 70, respectively. The data shows that 11/12 (91.67\%) of \textit{Confusing contracts} are destroyed by the developers and replaced by deploying a new one. However, only 19/70 (27.14\%) of \textit{Unmatched ERC20 Tokens} contracts are destroyed and replaced with a new one. For this inconsistency, developers might assume \textit{Confusing contracts} are more harmful than \textit{Unmatched ERC20 Tokens}. Therefore, when developers find a \textit{Confusing contracts} issue, they choose to destroy the contract and deploy a new one. For \textit{Unmatched ERC20 Tokens}, developers might still choose to use the contract, and destroy the contracts once the life cycle is finished. However, we cannot find their successor contracts. 

\subsubsection{Self-destructed Contracts on Blockchain vs. Etherscan.}
\label{sec:inconsist2}
We summarized 5 reasons why smart contracts were destructed based on 756 self-destructed contracts collected from Etherscan. However, there are millions of destructed contracts on Ethereum blockchain. The inconsistency between the number of self-destructed Contracts on Ethereum blockchain and Etherscan might make our finding not so reliable. To investigate this inconsistency, we collected all self-destructed-related traces by Jan. 2019 (The date is the same as the verified contract we collected from Etherscan.)  Table~\ref{tab:selfBC} shows the self-destructed-related information we collected from Ethereum blockchain. There are 30,486,241 self-destructed traces which were generated by 2,084,841 self-destructed contracts. When a contract is destructed, the contract will transfer its balance to another account; we call these contracts used to receive the balance as ``dest" accounts. We found 19,131,801 dest accounts. Noticed that one transaction can have several self-destructed-related traces and one self-destructed-related trace related to one self-destructed operation. There is a difference between the number of self-destruct traces and the number of self-destructed contracts because a contract that has been marked as \textit{destructed} still exists until the end of the transaction. The contracts can still be called and may execute further self-destructs. This mechanism was used by attackers to launch the DDoS attack in block No. 2.3M to 2.7M~\cite{greene2018investigation}. 27,560,501 (90.4\%) self-destruct traces and 19,127,397 (99.98\%) dest accounts were generated in this DDoS attack. Thus, to make the analysis more accurate, we removed the self-destructed-related information. generated on this DDoS attack, and there are 2,925,740 self-destruct traces generated by 2,080,319 self-destructed contracts. An interesting finding is that all the 2,080,319 self-destructed contracts are related to only 4,404 dest accounts. Among these 4,404 dest accounts, 2,716 are EOAs, 1,263 are self-destructed contracts and 425 are alived contracts (contracts that can be found on blockchain), while the number of dest accounts of 756 destructed contracts collected from Etherscan is 472 (415 EOA, 16 self-destructed contracts and 41 alived contracts). It shows that although there are large number of self-destructed contracts on Ethereum blockchain, most of them were generated by a limited number of accounts.

Previous work~\cite{di2020characterizing} also investigated the self-destructed contracts on Ethereum. They analyzed 7.3 M self-destructed contracts, and 7.2M (98.6\%) were short-lived contracts (4.2M, 57.53\%)~\footnote{A contract called short-lived when the creation and subsequent selfdestruct is executed in the same transaction.},  GasTokens (2.8M, 38.35\%), and ENS (Ethereum Name Service) deeds (0.2M, 2.74\%). All these 98.6\% self-destructed contracts will not be included in the analysis of these papers because these contracts usually do not contain much information. For example, the recommended template of recommended GasToken~\cite{gastoken} only contains seven instructions with two lines of code. Short-lived contracts are usually contract-created simple contracts that are created and destructed in a single transaction.  Thus, there is no need to open source these contracts on Etherscan and will not be included in our analysis. 

Our analysis and Di Angelo and Salzer's~\cite{di2020characterizing} finding can show that the self-destructed contracts we collected from Etherscan have a good coverage of selfdestruct function usages apart from DDoS attack, GasToken, and ENS deeds. It is also likely that we might miss reporting some usages because of the limitation of manual analysis used in this article. However,this article also highlights a new direction and might trigger more studies in the future.

\begin{table}
	\footnotesize
	\caption{self-destructed-related Info. on Ethereum blockchain} 
	\label{tab:selfBC}
	\centering
	%\begin{tabular}{c | c}
	\begin{tabular}{c | r  | r | r | r}
		\hline
		 & Trace & self-destructed Contracts & Dest Accounts & Trans.\\
		\hline
		Total &  30,486,241 &  2,084,841 & 19,131,801 & 374,735\\
		\hline
		Block No. 2.3M - 2.7M & 27,560,501 & 4,522 & 19,127,397 & 53,374 \\
		\hline
		Others & 2,925,740 & 2,080,319 & 4,404 & 321,361\\
	\end{tabular}	
\end{table}

\begin{table}
	\footnotesize
	\caption{Open Interview Questions (Excerpt)} 
	\label{tab:interview}
	\centering
	%\begin{tabular}{c | c}
	\begin{tabular}{c | l}
		\hline
		ID & Question\\
		\hline
		1 &  Do you used selfdestruct function?\\
		\hline
		2& Do you know the gas refund feature?\\
		\hline
		3 & Do you know selfdestruct function can refund gas?\\
		\hline
		4 & Do you add selfdestruct function for refunding gas? why?\\
		\hline
		5 & Do you know GasToken and how do you think about it?\\
	\end{tabular}	
\end{table}

\subsubsection{\textit{GasToken} Contracts on Blockchain vs. Survey Results.}
Previous work~\cite{di2020characterizing} reported that about 38.35\% self-destructed contracts on Ethereum blockchain are GasToken. In our survey, only one respondent mentioned that the motivation of including the \textit{selfdestruct} function is refunding gas. To investigate this inconsistency, we performed an interview to collect developers' perspectives. From our survey, 33 respondents claim that they will add the selfdestruct function in their smart contracts. 18 out of 33 left their emails, and we sent emails to ask them whether they agreed to have a further interview about the gas refund feature. Finally, 8 respondents accepted the interview invitations, and all interviews were performed remotely via Skype or WeChat. We conducted semi-structured interviews followed by Zhou et al.’s method~\cite{zou2019smart}. Specifically, we first introduced our work and asked the demographic questions shown in Section 3.2.2. The interviewees had an average experience of 3.9 years in smart contract developments with various roles, including 3 developers, 1 manager, 1 tester and 2 researchers. Then, we used some open questions listed in Table~\ref{tab:interview} to guide the discussion. The first question is used to confirm their qualification and all interviewees answered ``\textit{Yes}". For the second question, 6 interviewees said they did not know it, and the interviews were finished. For the other 2 interviewees, both said they had a deep understanding of the gas refund feature and knew the selfdestruct function could refund gas. Thus, we continued to the fourth and fifth questions to collect their feedback. As our interviews were semi-structured, we also asked follow-up questions to dig deeper according to their answers. For example, one interviewee said they have no need to use selfdestruct to return gas back, and we then asked them why this was the case. 

We found the following reasons why they will not include \textit{selfdestruct} function just for refunding gas. First, the most important reason is the lack of knowledge about this feature. Although the interviewees had an average experience of 3.9 years in smart contract developments, only two of them know this feature. Second, one interviewee said they focus more on the functionalities and the security of smart contracts. They do not pay for the gas and thus they do not care how to save gas for the contracts. Besides, both two interviewees mentioned that adding \textit{selfdestruct} function will increase development cost. However, only up to half of the gas used by a transaction that calls the \textit{selfdestruct} function will be returned. They believe the loss outweighs the gain. 

In terms of the GasToken, the two interviewees gave various views, but all the views were negative. The first interviewee mentioned that he did not believe GasToken can save gas or used it to make profits. Developers can only save a little gas fee when the gas price on free time is several times higher than mint time. If the gas price on free time is similar or lower than mint time, using GasTokens will even cost more, as the creation of the GasTokens will also cost gas. Besides, the miner can only receive half of the gas fee because of the GasToken, and they might refuse to process the transaction. Thus, the usage of GasTokens seems narrow. Another interviewee said that GasToken is harmful for the Ethereum ecosystem, as it will create a large number of useless contracts on Ethereum. All of these useless contracts will be stored in nodes because of the distributed ledger nature of Ethereum, which wastes many storage and network resources. 

Except from the views we collected from the interview, in the previous subsection (Section~\ref{sec:inconsist2}), we found that all the 2,080,319 self-destructed contracts only related to 4,404 dest accounts. This also shows that although there are a large number of GasTokens on Ethereum, they are only used by a small number of users. It is an interesting topic to conduct a comprehensive investigation of the GasTokens, e.g., their main usage scenarios, but they are out-of-scope to this paper. Thus, we will conduct the investigation in our future work.

\subsection{ Threats to Validity}
\noindent \textbf{Internal Validity.} 
In RQ1, we sent our survey to 996 developers and received 88 responses. The response rate is 8.84\%. We used the feedback of these 88 responses to summarize key reasons why developers include or exclude \textit{selfdestruct} function in their contracts. Due to the limited number of feedback. There might however still be other reasons we did not cover in our survey. We collected all contributors emails from the top 100 most popular smart contract related projects. We also tried to make our survey as simple as possible and give 2 respondents \$50 Amazon gift card to increase the response rate. We finally obtained a 8.84\% response rate, which is also acceptable~\cite{xia2019practitioners}. 

In RQ2, we use \textsc{SmartEmbed} to compute the similarity between smart contracts. If the similarity of two contracts larger than 0.6, we think they are a predecessor contract and its successor contract. The similarity threshold can influence the manual effort we need to pay. If the similarity is too large, we might miss some predecessor contracts and their successor contract. Otherwise, if the similarity is too small, we need to pay more effort to distinguishing whether the two contracts are relevant or not. The similarity threshold used in the paper of \textsc{SmartEmbed} is 0.95, and it found few contracts are relevant if the similarity is lower than 0.7. To reduce the number of unidentified relevant contracts, we conservatively reduce the threshold to 0.6. We used \textit{Open Card Sorting} to find 5 common self-destruct reasons. Due to the limitation of our understanding of the smart contracts, we might miss some self-destruct reasons. To reduce the threat of human factors, we followed the process of card sorting strictly, and the developers all have rich experience (\textgreater 3 years) in smart contract related research. Researchers can also use the same method we proposed in RQ2 to find other self-destruct reasons in the future. 

In RQ3, we use contracts whose number of transactions are larger than 500, and balance is larger than 1 Ether as the ground truth. We regard these contracts having a low probability of having permission problems. However, it is still possible we might find some functions in this group that have permission problems, but we believe the number of these functions is small. 

We performed an interview to collect why developers do not include selfdestruct function for refunding gas. However, only 8 respondents accepted our interview, and 6 of them did not provide much information as they do not know the gas refund feature of Ethereum. Thus, it is likely that some points might be missing. Fortunately, we still obtain many reasonable feedbacks which can answer the inconsistency. We acknowledge that a comprehensive and convincing investigation about the GasTokens is an interesting topic, but it is not the core focus of this paper and needs further research efforts that we reserve for future work.  

\noindent \textbf{External Validity.} The smart contracts used in this paper were up to Jan. 2019.  \textit{Solidity}, the most popular programming language for the smart contract, is fast-growing. From Jan. 2019 to the time of writing, there are 11 versions updated and released. Many new features have been removed and added in these versions. Ethereum also might be updated in the future through a hard fork~\cite{hardfork}. In this case, the self-destruct reasons might be changed because of a major update to Ethereum and Solidity. Addressing this threat needs more research effort, but the method we proposed to find the self-destruct reasons is still working. 

In this paper, we first cluster contracts with their creator addresses. However, it is likely that some developers may use multiple addresses to deploy smart contracts. In this case, our method may fail in finding successor contracts of some self-destructed contracts, and our analysis may miss uncovering some reasons for the use of self-destruct. Because of the anonymity of Ethereum, it is difficult to cluster contracts with multiple creator addresses, even if they are owned by the same developer. Fortunately, from the collection of 756 self-destructed contracts that we analyzed, we find that for 436 of them we could find their successor contracts. In total we found 1513 \textit{\textless predecessor contract, successor contract\textgreater} pairs. We use open card sorting to summarize the reasons why contracts self-destructed (details see Section 4.2.3). There are two iterations in the open card sorting. In the first iteration, we use 20\% of the cards (one card is a \textit{\textless predecessor contract, successor contract\textgreater} pairs) and get all the reasons. No new reason was found by using the remaining 80\% of cards in the second iteration, which means 1513 pairs were enough to support the analysis work reported in this paper.

%Besides, we find 436 self-destruct contracts among 756 contracts have their successor contracts with 1513 \textit{\textless predecessor contract, successor contract\textgreater} pairs. The number of pairs also show a good coverage of our method. 
%It is difficult to ensure that all developers have a good background of Solidity programming. Some developers might not familiar with \textit{selfdestruct} function. Therefore, it is possible that some feedback might contain incorrect information. To reduce the influence of this situation.

	\section{Related Work}
\label{sec:related}
%\smartEmbeding

\textsc{SmartEmbed}~\cite{smartembed, gao2020checking} is the first tool that uses a clone detection method to detect bugs in smart contracts.  The tool contains a training phase and a prediction phase. In the training phase, their dataset contains two parts, i.e., source code database and bug database. The source code database consists of the source code of all the open source smart contracts in the Ethereum. The bug database records the bugs of each smart contract in a source code database. \textsc{SmartEmbed} first converts each smart contract to an AST(abstract syntax tree). After normalizing the parameters and irrelevant information on the AST, \textsc{SmartEmbed} transfers the tree structure to a sequence representation. Then, they use \textit{Fasttext}~\cite{bojanowski2017enriching} to transfer code to embedding matrices. Finally, they compute the similarity between the given smart contracts with contracts in their database to find the clone contracts and clone related bugs.  Although the tool is aimed at finding bugs, their first step is computing the code similarity between the given smart contract and history contracts in their database. Therefore, we can modify their code to compute the similarity between two given smart contracts (used in RQ2).   

%ponzi
Bartoletti et al.~\cite{ponzi-origin} found that the infamous Ponzi schemes migrated to the digital world. Many frauds use Ethereum to design Ponzi schemes contracts for earning money. They manually analyzed 1,382 verified smart contracts on Etherscan and find 137 of them are Ponzi scheme contracts. Then, they divided these Ponzi scheme contracts into four categories, i.e., array-based pyramid schemes, tree-based pyramid schemes, handover schemes, and waterfall schemes. Bartoletti et al. opened their dataset to the public but do not provide a tool to detect whether a contract is a Ponzi scheme contract. To address this limitation, Chen et al.~\cite{ponzi-www} proposed a method that uses a machine learning algorithm (XGBoost~\cite{xgboost}) to distinguish Ponzi scheme contracts. They use account features and code features to train the module. The account features are extracted from the transactions, e.g., the number of payment transactions, the balance in the contracts. The code features can be obtained from contract bytecode. They count the frequency of each opcode in the contract bytecode. Both account features and code features do not need the source code of contracts. Therefore, their method can predict arbitrary contracts on the Ethereum. 

%Bartoletti et al.~\cite{ponzi-origin} found that many frauds use Ethereum to design Ponzi schemes contracts for earning money. They manually analysis 1382 verified smart contracts on Etherscan and find 137 of them are Ponzi scheme contracts. Then, they divided these Ponzi scheme contracts into four categories, i.e., array-based pyramid schemes, tree-based pyramid schemes, handover schemes, and waterfall schemes. Bartoletti et al. opened their dataset to the public but do not provide a tool to detect whether a contract is a Ponzi scheme contract. To address this limitation, Chen et al.~\cite{ponzi-www} proposed a method that use a machine learning algorithm (XGBoost~\cite{xgboost}) to distinguish Ponzi scheme contracts. They use account features and code features to train the module. The account features are extracted from the transactions and the code features can be obtained from contract bytecode. Both account features and code features do not need the source code of contracts. Therefore, their method can predict arbitrary contracts on the Ethereum. 

%oyente
Oyente~\cite{oyente} is the first tool for security examination for smart contracts based on symbolic execution. Their work introduces four security issues on smart contracts, i.e., mishandled exception, transaction-ordering dependence, timestamps dependence, and re-entrancy attack. To detect these security issues, Oyente first constructs a CFG (control flow graph) based on symbolic execution. After that, they design different rules to detect these four security issues. Kalra et al.~\cite{Zeus} proposed a tool named Zeus, which can detect seven kinds of security problems; four of them are the same with Oyente; the other three issues are \textit{failed send, interger overflow/underflow} and \textit{transaction state dependence}. Zeus can detect security issues at the source code level. They use LLVM bytecode to represent the Solidity source and detect related patterns through LLVM bytecode. ContractFuzzer~\cite{Contractfuzzer} is the first fuzzer to detect seven security issues in smart contracts. Four security issues are the same as Oyente; the other three issues are \textit{gasless send}, \textit{dangerous delegatecall} and \textit{freezing ether}. ContractFuzzer utilizes ABI (abstract binary interface) of smart contracts to generate fuzzing inputs and defines test oracles to detect security issues. 

%Chen et al.~\cite{domainSmells} define 20 domain-specific code smells for smart contracts and divide them into three categories, i.e., \textit{security smells}, \textit{architecture smells}, and \textit{usability smells}. They assign an impact level for each code smells. Smells with impact level 1-3 can lead to unwanted behaviors of contract, e.g., crashing or being attacked.  \vspace{-0.1cm}

Chen et al.~\cite{chen2020defining} define 20 smart contract defects on Ethereum. They first crawl 17,128 Stack Exchange posts and use key words to filter solidity related posts. After getting Solidity related posts, they use \textit{Open Card Sorting} to find 20 contract defects and divide them into five categories, i.e., \textit{security, availability, performance, maintainability}, and \textit{reusability defects}. According to their paper, although previous works define several security defects, they did not consider the practitioners' perspective. Therefore, they design an online survey to collect feedback from developers. The feedback shows that all the defined contract defects are harmful to smart contracts. They assign five impact levels the defined 20 contract defects. Defects with impact level 1-3 can lead to unwanted behaviors of contract, e.g., crashing or a contract being attacked.

Li et al.~\cite{li2018detecting} proposed a symbolic execution analysis tool named SOLAR to detect violations of  two standards, i.e., ERC20 and ERC721. SOLAR is built on top of Manticore~\cite{mossberg2019manticore}  which is a well-known symbolic execution framework for smart contracts, and uses boolector~\cite{niemetz2014boolector} as the SMT solver to check the symbolic constraints. Since Manticore does not fully support EVM instructions, SOLAR is extended to fully support EVM instructions. Their experimental results show that SOLAR is significantly more effective than previous tools, e.g, Mythril~\cite{Mythril}, and can find more errors with fewer false positives.  

Di Angelo and Salzer.~\cite{di2020characterizing} introduced the usage of smart contracts on Ethereum by analyzing about 20 million deployed smart contracts. They defined ten kinds of usages of smart contracts on Ethereum and found that most of the deployed smart contracts remain unused and tokens are the most popular applications of Ethereum. self-destructed contracts is one important research dimension in their work. They found 7.3M self-destructed contracts on their dataset. According to their analysis, 4.2 M of them were created and then self-destructed in the same transaction; 2.8 M were GasTokens, and 0.2 M were ENS (Ethereum Name Service) deeds. Eight thousand of the remaining nine thousand contracts were self-destructed for unknown reasons and 778 were wallets. Their other work ~\cite{di2019mayflies} also investigated the usage of self-destructed contracts and found 48,506 contracts self-destructed multiple times (some of them up to 10 920 times). They called a contract a mayfly if the contract was created and self-destructed in the same transaction. They found 1,856,655 mayflies that were created by just 8,992 distinct addresses, and most of the mayflies appeared during the DDos period of 2016. All of their work and this work investigated the usage of selfdestruct function on Ethereum, but our work is more comprehensive and has a different focus. Specifically, we first conducted an online survey to collect developers' feedback about why they add or do not add selfdestruct function. Then, we proposed a method to find the reasons why smart contracts self-destructed. Finally, we propose a tool to detect two problems that might shorten the lifespan of smart contracts.

	\section{Conclusions and Future Work}
In this paper, we conducted a comprehensive empirical study on the use of the \textit{selfdestruct} function on Ethereum. To understand the smart contract developers' perspective, in RQ1, we designed an online survey to collect reasons from developers why they include and exclude the \textit{selfdestruct} function in their contracts. We summarized 6 reasons for including and  6 reasons for  excluding \textit{selfdestruct} function in their contracts, respectively. The feedback also shows that 22 / 33 respondents claim that they add \textit{selfdestruct} function for security concerns or to upgrade contracts. These two motivations can lead to redeployment of smart contracts after developers destruct the contracts.  According to this information, we propose an approach that can find the upgrade version of the self-destructed contracts in RQ2.  After that, we used the \textit{open card sorting} method and summarized 5 reasons why contracts might destruct. Two of them -- \textit{Unmatched ERC20 Token} and \textit{Limits of Permission} -- can affect the life span of smart contracts. To detect these problems, we developed a new tool named \textsc{LifeScope} in RQ3, which reports 0 false positive / negative in detecting \textit{Unmatched ERC20 Token}, and achieves an  F-measure and AUC of 77.89\% and 0.8673 for detecting the \textit{Limits of Permission} issue. Finally, to help developers use the \textit{selfdestruct} function better, we give 6 suggestions based on the feedback of our survey and our smart contract analysis.

%In the future, we plan to summarize more self-destruct reasons of smart contracts in Ethereum, as more contracts will be deployed, and new features will be added in Solidity / Ethereum. We will update \textsc{LifeScope} to detect the additional problems that lead to the self-destruct of contracts. 

Apart from the \textit{selfdestruct} function, using \textit{Delegatecall} or \textit{Callcode} can also be used to design an upgradeable smart contract~\cite{upgradable}. Specifically, we need a proxy contract and a logic contract to design an upgradeable contract. The proxy contract stores the storage variables and Ethers, and the logic contract contains the logical code. The proxy contract uses \textit{Delegatecall} to call the code of the logic contract. If there are some security issues found on the logic contract, developers can deploy a new logic contract and discard the old logic contract. By comparing the difference between the old logic contract and new logic contract, we can find some reasons why contracts need upgrade. In Section 6.3, we only investigate the transaction of destructed unmatched ERC20 tokens. However, it is likely that the matched ERC20 tokens also do not transfer values back to the users. Also, the upgradeable contracts will also lock the tokens of users. In the future, we will conduct more comprehensive work to investigate the value lock on ERC20 tokens. 

We found 2,789 (5.1\%) smart contracts in total contain the selfdestruct function, while only 199 (0.36\%) contracts contain the Delegatecall or Callcode functions. The data size is one of the main reasons why we choose to investigate \textit{selfdestruct} function first. In the future, we plan to conduct an empirical study to investigate the upgradeable smart contracts on Ethereum when there are more smart contracts that contain Delegatecall or Callcode function. Specifically, there are many new issues of designing an upgradeable smart contract. For example, although the old logic contract is discarded, the contract can still be called by other contracts, which might lead to new security issues; designing an upgradeable smart contract can also increase the development cost~\cite{chen2020maintaining}. Thus, we first plan to investigate developers’ motivation about why they design their contracts as upgradeable. Then, we plan to design a method to find the \textit{\textless old logic contract, new logic contract\textgreater} pairs. Unlike \textit{selfdestruct} function where we need to compare similarity to find the pair, for the upgradeable contract, we can check the transaction details of the proxy contract. From the transactions, we can find the discarded logic contract and new logic contract. After that, we can compare their differences to find the reasons why developers upgrade a smart contract. Finally, we will update \textsc{LifeScope} to detect the additional problems. 

We analyzed self-destructed contracts with source code and conducted a preliminary analysis for the self-destructed contracts on Ethereum blockchain. In the future, we will perform a comprehensive empirical study to analyze the self-destructed contract on Ethereum. Specifically, we first investigate how many GasTokens on Ethereum; how many users of the GasTokens, and the usage scenarios of the GasTokens. Besides, it is interesting to analyze remaining self-destructed contracts. For example, are there any other DDoS attacks happening; why millions of self-destructed contracts are only related to 4,404 dest accounts, and why some developers created a large number of self-destructed contracts. 

We observed the difference between a PS pair to summarize the reasons why smart contracts were destructed. However, we only stand at a high level to present our observations instead of digging out a more detailed reason. Specifically, \textit{Setting Changes} and \textit{Functionality Changes} are two broad definitions. Developers might change the functionalities for several reasons, e.g., adding business requirements, fixing bugs, increasing readabilities. Besides, when multiple changes happen, we regard all of them as the reasons that contributed to the self-destruct. Actually, their importance might be different, and maybe only one of them was the real reason why smart contracts were destructed. In the future, we will conduct a more comprehensive empirical study to find more specific reasons that lead a contract self-destruct.

\noindent{\textbf{Acknowledgment. }} This research/project is supported by ARC Laureate Fellowship FL190100035, and the National Research Foundation, Singapore under its Industry Alignment Fund – Pre-positioning (IAF-PP) Funding Initiative. Any opinions, findings and conclusions or recommendations expressed in this material are those of the author(s) and do not reflect the views of National Research Foundation, Singapore.
	
\balance
	\bibliographystyle{ACM-Reference-Format}
	\bibliography{ref}


\begin{thebibliography}{72}

%%% the Web address.

\ifx \showCODEN    \undefined \def \showCODEN     #1{\unskip}     \fi
\ifx \showDOI      \undefined \def \showDOI       #1{#1}\fi
\ifx \showISBNx    \undefined \def \showISBNx     #1{\unskip}     \fi
\ifx \showISBNxiii \undefined \def \showISBNxiii  #1{\unskip}     \fi
\ifx \showISSN     \undefined \def \showISSN      #1{\unskip}     \fi
\ifx \showLCCN     \undefined \def \showLCCN      #1{\unskip}     \fi
\ifx \shownote     \undefined \def \shownote      #1{#1}          \fi
\ifx \showarticletitle \undefined \def \showarticletitle #1{#1}   \fi
\ifx \showURL      \undefined \def \showURL       {\relax}        \fi
% invisible to TeX
\providecommand\bibfield[2]{#2}
\providecommand\bibinfo[2]{#2}
\providecommand\natexlab[1]{#1}
\providecommand\showeprint[2][]{arXiv:#2}

\bibitem[\protect\citeauthoryear{??}{DAO}{2018}]%
        {DAOAttack}
 \bibinfo{year}{Apr., 2018}\natexlab{}.
\newblock \bibinfo{booktitle}{\emph{Understanding The DAO Attack}}.
\newblock
\urldef\tempurl%
\url{https://www.coindesk.com/understanding-dao-hack-journalists/}
\showURL{%
\tempurl}


\bibitem[\protect\citeauthoryear{??}{dap}{2019}]%
        {dapp}
 \bibinfo{year}{Apr., 2019}\natexlab{}.
\newblock \bibinfo{booktitle}{\emph{Decentralized application}}.
\newblock
\urldef\tempurl%
\url{https://en.wikipedia.org/wiki/Decentralized\_application}
\showURL{%
\tempurl}


\bibitem[\protect\citeauthoryear{??}{erc}{2018}]%
        {erc20}
 \bibinfo{year}{April., 2018}\natexlab{}.
\newblock \bibinfo{booktitle}{\emph{ERC20}}.
\newblock
\urldef\tempurl%
\url{https://github.com/ethereum/EIPs/blob/master/EIPS/eip-20.md}
\showURL{%
\tempurl}


\bibitem[\protect\citeauthoryear{??}{bes}{2019}]%
        {bestPractice}
 \bibinfo{year}{Aug., 2019}\natexlab{}.
\newblock \bibinfo{booktitle}{\emph{Ethereum Smart Contract Security Best
  Practices}}.
\newblock
\urldef\tempurl%
\url{https://consensys.github.io/smart-contract-best-practices/}
\showURL{%
\tempurl}


\bibitem[\protect\citeauthoryear{??}{Myt}{2019}]%
        {Mythril}
 \bibinfo{year}{Aug., 2019}\natexlab{}.
\newblock \bibinfo{booktitle}{\emph{Mythril: Security analysis tool for EVM
  bytecode.}}
\newblock
\urldef\tempurl%
\url{https://github.com/ConsenSys/mythril}
\showURL{%
\tempurl}


\bibitem[\protect\citeauthoryear{??}{rip}{2019}]%
        {ripple}
 \bibinfo{year}{Aug., 2019}\natexlab{}.
\newblock \bibinfo{booktitle}{\emph{Ripple Coin}}.
\newblock
\urldef\tempurl%
\url{https://ripple.com/}
\showURL{%
\tempurl}


\bibitem[\protect\citeauthoryear{??}{dif}{2020}]%
        {diffchecker}
 \bibinfo{year}{Aug., 2020}\natexlab{}.
\newblock \bibinfo{booktitle}{\emph{DiffChecker}}.
\newblock
\urldef\tempurl%
\url{https://github.com/trembacz/diff-checker}
\showURL{%
\tempurl}


\bibitem[\protect\citeauthoryear{??}{upg}{2020}]%
        {upgradable}
 \bibinfo{year}{Dec., 2020}\natexlab{}.
\newblock \bibinfo{booktitle}{\emph{EIP-2535: Diamond Standard}}.
\newblock
\urldef\tempurl%
\url{https://eips.ethereum.org/EIPS/eip-2535}
\showURL{%
\tempurl}


\bibitem[\protect\citeauthoryear{??}{cry}{2019}]%
        {cryptokitties}
 \bibinfo{year}{Feb., 2019}\natexlab{}.
\newblock \bibinfo{booktitle}{\emph{Cryptokitties}}.
\newblock
\urldef\tempurl%
\url{https://www.cryptokitties.co/}
\showURL{%
\tempurl}


\bibitem[\protect\citeauthoryear{??}{ETC}{2019}]%
        {ETC}
 \bibinfo{year}{Feb., 2019}\natexlab{}.
\newblock \bibinfo{booktitle}{\emph{{Ethereum Classic}}}.
\newblock
\urldef\tempurl%
\url{https://en.wikipedia.org/wiki/Ethereum_Classic}
\showURL{%
\tempurl}


\bibitem[\protect\citeauthoryear{??}{Exp}{2020}]%
        {Expanse}
 \bibinfo{year}{Feb., 2020}\natexlab{}.
\newblock \bibinfo{booktitle}{\emph{{Expanse}}}.
\newblock
\urldef\tempurl%
\url{https://coinswitch.co/info/expanse/what-is-expanse}
\showURL{%
\tempurl}


\bibitem[\protect\citeauthoryear{??}{blo}{2019}]%
        {blockchain-wiki}
 \bibinfo{year}{Jan., 2019}\natexlab{}.
\newblock \bibinfo{booktitle}{\emph{Blockchain}}.
\newblock
\urldef\tempurl%
\url{https://en.wikipedia.org/wiki/Blockchain}
\showURL{%
\tempurl}


\bibitem[\protect\citeauthoryear{??}{Cry}{2019}]%
        {Cryptocurrency-wiki}
 \bibinfo{year}{Jan., 2019}\natexlab{}.
\newblock \bibinfo{booktitle}{\emph{Cryptocurrency}}.
\newblock
\urldef\tempurl%
\url{https://en.wikipedia.org/wiki/Cryptocurrency}
\showURL{%
\tempurl}


\bibitem[\protect\citeauthoryear{??}{eth}{2019}]%
        {ethereum}
 \bibinfo{year}{Jan., 2019}\natexlab{}.
\newblock \bibinfo{booktitle}{\emph{Ethereum.org}}.
\newblock
\urldef\tempurl%
\url{https://www.ethereum.org/}
\showURL{%
\tempurl}


\bibitem[\protect\citeauthoryear{??}{gas}{2021}]%
        {gastoken}
 \bibinfo{year}{June., 2021}\natexlab{}.
\newblock \bibinfo{booktitle}{\emph{GasToken.io}}.
\newblock
\urldef\tempurl%
\url{https://gastoken.io/}
\showURL{%
\tempurl}


\bibitem[\protect\citeauthoryear{??}{Eth}{2018}]%
        {EtherScan}
 \bibinfo{year}{Mar., 2018}\natexlab{}.
\newblock \bibinfo{booktitle}{\emph{EtherScan}}.
\newblock
\urldef\tempurl%
\url{https://etherscan.io/}
\showURL{%
\tempurl}


\bibitem[\protect\citeauthoryear{??}{sol}{2018}]%
        {solc}
 \bibinfo{year}{Mar., 2018}\natexlab{}.
\newblock \bibinfo{booktitle}{\emph{The Solidity Contract-Oriented Programming
  Language}}.
\newblock
\urldef\tempurl%
\url{https://github.com/ethereum/solidity}
\showURL{%
\tempurl}


\bibitem[\protect\citeauthoryear{??}{Sol}{2018}]%
        {Solidity}
 \bibinfo{year}{Mar., 2018}\natexlab{}.
\newblock \bibinfo{booktitle}{\emph{Solidity Document}}.
\newblock
\urldef\tempurl%
\url{http://solidity.readthedocs.io}
\showURL{%
\tempurl}


\bibitem[\protect\citeauthoryear{??}{har}{2019}]%
        {hardfork}
 \bibinfo{year}{Oct., 2019}\natexlab{}.
\newblock \bibinfo{booktitle}{\emph{Blockchain Hard Fork}}.
\newblock
\urldef\tempurl%
\url{https://www.investopedia.com/terms/h/hard-fork.asp}
\showURL{%
\tempurl}


\bibitem[\protect\citeauthoryear{??}{Cam}{2019}]%
        {CamelCase}
 \bibinfo{year}{Sept., 2019}\natexlab{}.
\newblock \bibinfo{booktitle}{\emph{Camel Case}}.
\newblock
\urldef\tempurl%
\url{https://en.wikipedia.org/wiki/Camel_case/}
\showURL{%
\tempurl}


\bibitem[\protect\citeauthoryear{Baeza-Yates, Ribeiro-Neto,
  et~al\mbox{.}}{Baeza-Yates et~al\mbox{.}}{1999}]%
        {tfidf}
\bibfield{author}{\bibinfo{person}{Ricardo Baeza-Yates},
  \bibinfo{person}{Berthier Ribeiro-Neto}, {et~al\mbox{.}}}
  \bibinfo{year}{1999}\natexlab{}.
\newblock \bibinfo{booktitle}{\emph{Modern information retrieval}}.
  Vol.~\bibinfo{volume}{463}.
\newblock \bibinfo{publisher}{ACM press New York}.
\newblock


\bibitem[\protect\citeauthoryear{Bartoletti, Carta, Cimoli, and
  Saia}{Bartoletti et~al\mbox{.}}{2020}]%
        {ponzi-origin}
\bibfield{author}{\bibinfo{person}{Massimo Bartoletti},
  \bibinfo{person}{Salvatore Carta}, \bibinfo{person}{Tiziana Cimoli}, {and}
  \bibinfo{person}{Roberto Saia}.} \bibinfo{year}{2020}\natexlab{}.
\newblock \showarticletitle{Dissecting Ponzi schemes on Ethereum:
  identification, analysis, and impact}.
\newblock \bibinfo{journal}{\emph{Future Generation Computer Systems}}
  \bibinfo{volume}{102} (\bibinfo{year}{2020}), \bibinfo{pages}{259--277}.
\newblock


\bibitem[\protect\citeauthoryear{Bojanowski, Grave, Joulin, and
  Mikolov}{Bojanowski et~al\mbox{.}}{2017}]%
        {bojanowski2017enriching}
\bibfield{author}{\bibinfo{person}{Piotr Bojanowski}, \bibinfo{person}{Edouard
  Grave}, \bibinfo{person}{Armand Joulin}, {and} \bibinfo{person}{Tomas
  Mikolov}.} \bibinfo{year}{2017}\natexlab{}.
\newblock \showarticletitle{Enriching word vectors with subword information}.
\newblock \bibinfo{journal}{\emph{Transactions of the Association for
  Computational Linguistics}}  \bibinfo{volume}{5} (\bibinfo{year}{2017}),
  \bibinfo{pages}{135--146}.
\newblock


\bibitem[\protect\citeauthoryear{Bosu, Iqbal, Shahriyar, and Chakraborty}{Bosu
  et~al\mbox{.}}{2019}]%
        {bosu2019understanding}
\bibfield{author}{\bibinfo{person}{Amiangshu Bosu}, \bibinfo{person}{Anindya
  Iqbal}, \bibinfo{person}{Rifat Shahriyar}, {and} \bibinfo{person}{Partha
  Chakraborty}.} \bibinfo{year}{2019}\natexlab{}.
\newblock \showarticletitle{Understanding the motivations, challenges and needs
  of blockchain software developers: A survey}.
\newblock \bibinfo{journal}{\emph{Empirical Software Engineering}}
  \bibinfo{volume}{24}, \bibinfo{number}{4} (\bibinfo{year}{2019}),
  \bibinfo{pages}{2636--2673}.
\newblock


\bibitem[\protect\citeauthoryear{Chakraborty, Shahriyar, Iqbal, and
  Bosu}{Chakraborty et~al\mbox{.}}{2018}]%
        {chakraborty2018understanding}
\bibfield{author}{\bibinfo{person}{Partha Chakraborty}, \bibinfo{person}{Rifat
  Shahriyar}, \bibinfo{person}{Anindya Iqbal}, {and} \bibinfo{person}{Amiangshu
  Bosu}.} \bibinfo{year}{2018}\natexlab{}.
\newblock \showarticletitle{Understanding the software development practices of
  blockchain projects: a survey}. In \bibinfo{booktitle}{\emph{Proceedings of
  the 12th ACM/IEEE International Symposium on Empirical Software Engineering
  and Measurement}}. \bibinfo{pages}{1--10}.
\newblock


\bibitem[\protect\citeauthoryear{Chen, Xia, Lo, Grundy, Luo, and Chen}{Chen
  et~al\mbox{.}}{2020c}]%
        {chen2020defining}
\bibfield{author}{\bibinfo{person}{Jiachi Chen}, \bibinfo{person}{Xin Xia},
  \bibinfo{person}{David Lo}, \bibinfo{person}{John Grundy},
  \bibinfo{person}{Xiapu Luo}, {and} \bibinfo{person}{Ting Chen}.}
  \bibinfo{year}{2020}\natexlab{c}.
\newblock \showarticletitle{Defining Smart Contract Defects on Ethereum}.
\newblock \bibinfo{journal}{\emph{IEEE Transactions on Software Engineering}}
  (\bibinfo{year}{2020}).
\newblock


\bibitem[\protect\citeauthoryear{Chen, Xia, Lo, Grundy, and Yang}{Chen
  et~al\mbox{.}}{2020b}]%
        {chen2020maintaining}
\bibfield{author}{\bibinfo{person}{Jiachi Chen}, \bibinfo{person}{Xin Xia},
  \bibinfo{person}{David Lo}, \bibinfo{person}{John Grundy}, {and}
  \bibinfo{person}{Xiaohu Yang}.} \bibinfo{year}{2020}\natexlab{b}.
\newblock \showarticletitle{Maintaining Smart Contracts on Ethereum: Issues,
  Techniques, and Future Challenges}.
\newblock \bibinfo{journal}{\emph{arXiv preprint arXiv:2007.00286}}
  (\bibinfo{year}{2020}).
\newblock


\bibitem[\protect\citeauthoryear{Chen, Cao, Li, Luo, Gu, Zhang, Liao, Zhu,
  Chen, He, et~al\mbox{.}}{Chen et~al\mbox{.}}{2020a}]%
        {chen2020soda}
\bibfield{author}{\bibinfo{person}{Ting Chen}, \bibinfo{person}{Rong Cao},
  \bibinfo{person}{Ting Li}, \bibinfo{person}{Xiapu Luo},
  \bibinfo{person}{Guofei Gu}, \bibinfo{person}{Yufei Zhang},
  \bibinfo{person}{Zhou Liao}, \bibinfo{person}{Hang Zhu},
  \bibinfo{person}{Gang Chen}, \bibinfo{person}{Zheyuan He}, {et~al\mbox{.}}}
  \bibinfo{year}{2020}\natexlab{a}.
\newblock \showarticletitle{SODA: A Generic Online Detection Framework for
  Smart Contracts}. In \bibinfo{booktitle}{\emph{Proceedings of the 27th
  Network and Distributed System Security Symposium}}.
\newblock


\bibitem[\protect\citeauthoryear{Chen and Guestrin}{Chen and Guestrin}{2016}]%
        {xgboost}
\bibfield{author}{\bibinfo{person}{Tianqi Chen} {and} \bibinfo{person}{Carlos
  Guestrin}.} \bibinfo{year}{2016}\natexlab{}.
\newblock \showarticletitle{Xgboost: A scalable tree boosting system}. In
  \bibinfo{booktitle}{\emph{Proceedings of the 22nd acm sigkdd international
  conference on knowledge discovery and data mining}}. ACM,
  \bibinfo{pages}{785--794}.
\newblock


\bibitem[\protect\citeauthoryear{Chen, Zhang, Li, Luo, Wang, Cao, Xiao, and
  Zhang}{Chen et~al\mbox{.}}{2019}]%
        {chen2019tokenscope}
\bibfield{author}{\bibinfo{person}{Ting Chen}, \bibinfo{person}{Yufei Zhang},
  \bibinfo{person}{Zihao Li}, \bibinfo{person}{Xiapu Luo},
  \bibinfo{person}{Ting Wang}, \bibinfo{person}{Rong Cao},
  \bibinfo{person}{Xiuzhuo Xiao}, {and} \bibinfo{person}{Xiaosong Zhang}.}
  \bibinfo{year}{2019}\natexlab{}.
\newblock \showarticletitle{TokenScope: Automatically detecting inconsistent
  behaviors of cryptocurrency tokens in Ethereum}. In
  \bibinfo{booktitle}{\emph{Proceedings of the 2019 ACM SIGSAC Conference on
  Computer and Communications Security}}. \bibinfo{pages}{1503--1520}.
\newblock


\bibitem[\protect\citeauthoryear{Chen, Zhu, Li, Chen, Li, Luo, Lin, and
  Zhange}{Chen et~al\mbox{.}}{2018}]%
        {chen2018infocom}
\bibfield{author}{\bibinfo{person}{Ting Chen}, \bibinfo{person}{Yuxiao Zhu},
  \bibinfo{person}{Zihao Li}, \bibinfo{person}{Jiachi Chen},
  \bibinfo{person}{Xiaoqi Li}, \bibinfo{person}{Xiapu Luo},
  \bibinfo{person}{Xiaodong Lin}, {and} \bibinfo{person}{Xiaosong Zhange}.}
  \bibinfo{year}{2018}\natexlab{}.
\newblock \showarticletitle{Understanding ethereum via graph analysis}. In
  \bibinfo{booktitle}{\emph{IEEE INFOCOM 2018-IEEE Conference on Computer
  Communications}}. IEEE, \bibinfo{pages}{1484--1492}.
\newblock


\bibitem[\protect\citeauthoryear{Cohen}{Cohen}{1960}]%
        {kappa}
\bibfield{author}{\bibinfo{person}{Jacob Cohen}.}
  \bibinfo{year}{1960}\natexlab{}.
\newblock \showarticletitle{A coefficient of agreement for nominal scales}.
\newblock \bibinfo{journal}{\emph{Educational and psychological measurement}}
  \bibinfo{volume}{20}, \bibinfo{number}{1} (\bibinfo{year}{1960}),
  \bibinfo{pages}{37--46}.
\newblock


\bibitem[\protect\citeauthoryear{Di~Angelo and Salzer}{Di~Angelo and
  Salzer}{2019}]%
        {di2019mayflies}
\bibfield{author}{\bibinfo{person}{Monika Di~Angelo} {and}
  \bibinfo{person}{Gernot Salzer}.} \bibinfo{year}{2019}\natexlab{}.
\newblock \showarticletitle{Mayflies, breeders, and busy bees in Ethereum:
  smart contracts over time}. In \bibinfo{booktitle}{\emph{Proceedings of the
  Third ACM Workshop on Blockchains, Cryptocurrencies and Contracts}}.
  \bibinfo{pages}{1--10}.
\newblock


\bibitem[\protect\citeauthoryear{Di~Angelo and Salzer}{Di~Angelo and
  Salzer}{2020}]%
        {di2020characterizing}
\bibfield{author}{\bibinfo{person}{Monika Di~Angelo} {and}
  \bibinfo{person}{Gernot Salzer}.} \bibinfo{year}{2020}\natexlab{}.
\newblock \showarticletitle{Characterizing types of smart contracts in the
  ethereum landscape}. In \bibinfo{booktitle}{\emph{International Conference on
  Financial Cryptography and Data Security}}. Springer,
  \bibinfo{pages}{389--404}.
\newblock


\bibitem[\protect\citeauthoryear{Efanov and Roschin}{Efanov and
  Roschin}{2018}]%
        {efanov2018all}
\bibfield{author}{\bibinfo{person}{Dmitry Efanov} {and} \bibinfo{person}{Pavel
  Roschin}.} \bibinfo{year}{2018}\natexlab{}.
\newblock \showarticletitle{The all-pervasiveness of the blockchain
  technology}.
\newblock \bibinfo{journal}{\emph{Procedia Computer Science}}
  \bibinfo{volume}{123} (\bibinfo{year}{2018}), \bibinfo{pages}{116--121}.
\newblock


\bibitem[\protect\citeauthoryear{Fan, Xia, da~Costa, Lo, Hassan, and Li}{Fan
  et~al\mbox{.}}{2019}]%
        {fan2019impact}
\bibfield{author}{\bibinfo{person}{Yuanrui Fan}, \bibinfo{person}{Xin Xia},
  \bibinfo{person}{Daniel~Alencar da Costa}, \bibinfo{person}{David Lo},
  \bibinfo{person}{Ahmed~E Hassan}, {and} \bibinfo{person}{Shanping Li}.}
  \bibinfo{year}{2019}\natexlab{}.
\newblock \showarticletitle{The Impact of Changes Mislabeled by SZZ on
  Just-in-Time Defect Prediction}.
\newblock \bibinfo{journal}{\emph{IEEE Transactions on Software Engineering}}
  (\bibinfo{year}{2019}).
\newblock


\bibitem[\protect\citeauthoryear{Fan, Xia, Lo, and Hassan}{Fan
  et~al\mbox{.}}{2018}]%
        {fan2018chaff}
\bibfield{author}{\bibinfo{person}{Yuanrui Fan}, \bibinfo{person}{Xin Xia},
  \bibinfo{person}{David Lo}, {and} \bibinfo{person}{Ahmed~E Hassan}.}
  \bibinfo{year}{2018}\natexlab{}.
\newblock \showarticletitle{Chaff from the wheat: Characterizing and
  determining valid bug reports}.
\newblock \bibinfo{journal}{\emph{IEEE transactions on software engineering}}
  (\bibinfo{year}{2018}).
\newblock


\bibitem[\protect\citeauthoryear{Ferreira~Torres, Baden, Norvill, and
  Jonker}{Ferreira~Torres et~al\mbox{.}}{2019}]%
        {ferreira2019aegis}
\bibfield{author}{\bibinfo{person}{Christof Ferreira~Torres},
  \bibinfo{person}{Mathis Baden}, \bibinfo{person}{Robert Norvill}, {and}
  \bibinfo{person}{Hugo Jonker}.} \bibinfo{year}{2019}\natexlab{}.
\newblock \showarticletitle{{\AE}GIS: Smart Shielding of Smart Contracts}. In
  \bibinfo{booktitle}{\emph{Proceedings of the 2019 ACM SIGSAC Conference on
  Computer and Communications Security}}. \bibinfo{pages}{2589--2591}.
\newblock


\bibitem[\protect\citeauthoryear{Flouri, Giaquinta, Kobert, and Ukkonen}{Flouri
  et~al\mbox{.}}{2015}]%
        {LCS}
\bibfield{author}{\bibinfo{person}{Tom{\'a}s Flouri}, \bibinfo{person}{Emanuele
  Giaquinta}, \bibinfo{person}{Kassian Kobert}, {and} \bibinfo{person}{Esko
  Ukkonen}.} \bibinfo{year}{2015}\natexlab{}.
\newblock \showarticletitle{Longest common substrings with k mismatches}.
\newblock \bibinfo{journal}{\emph{Inform. Process. Lett.}}
  \bibinfo{volume}{115}, \bibinfo{number}{6-8} (\bibinfo{year}{2015}),
  \bibinfo{pages}{643--647}.
\newblock


\bibitem[\protect\citeauthoryear{Fowler and Beck}{Fowler and Beck}{1999}]%
        {smellDefinition}
\bibfield{author}{\bibinfo{person}{Martin Fowler} {and} \bibinfo{person}{Kent
  Beck}.} \bibinfo{year}{1999}\natexlab{}.
\newblock \bibinfo{booktitle}{\emph{Refactoring: improving the design of
  existing code}}.
\newblock \bibinfo{publisher}{Addison-Wesley Professional}.
\newblock


\bibitem[\protect\citeauthoryear{Fr{\"o}wis, Fuchs, and B{\"o}hme}{Fr{\"o}wis
  et~al\mbox{.}}{2019}]%
        {frowis2019detecting}
\bibfield{author}{\bibinfo{person}{Michael Fr{\"o}wis},
  \bibinfo{person}{Andreas Fuchs}, {and} \bibinfo{person}{Rainer B{\"o}hme}.}
  \bibinfo{year}{2019}\natexlab{}.
\newblock \showarticletitle{Detecting token systems on ethereum}. In
  \bibinfo{booktitle}{\emph{International conference on financial cryptography
  and data security}}. Springer, \bibinfo{pages}{93--112}.
\newblock


\bibitem[\protect\citeauthoryear{Gao, Jayasundara, Jiang, Xia, Lo, and
  Grundy}{Gao et~al\mbox{.}}{2019}]%
        {smartembed}
\bibfield{author}{\bibinfo{person}{Zhipeng Gao}, \bibinfo{person}{Vinoj
  Jayasundara}, \bibinfo{person}{Lingxiao Jiang}, \bibinfo{person}{Xin Xia},
  \bibinfo{person}{David Lo}, {and} \bibinfo{person}{John Grundy}.}
  \bibinfo{year}{2019}\natexlab{}.
\newblock \showarticletitle{SmartEmbed: A Tool for Clone and Bug Detection in
  Smart Contracts through Structural Code Embedding}.
\newblock \bibinfo{journal}{\emph{35th IEEE International Conference on
  Software Maintenance and Evolution (ICSME)}} (\bibinfo{year}{2019}).
\newblock


\bibitem[\protect\citeauthoryear{Gao, Jiang, Xia, Lo, and Grundy}{Gao
  et~al\mbox{.}}{2020}]%
        {gao2020checking}
\bibfield{author}{\bibinfo{person}{Zhipeng Gao}, \bibinfo{person}{Lingxiao
  Jiang}, \bibinfo{person}{Xin Xia}, \bibinfo{person}{David Lo}, {and}
  \bibinfo{person}{John Grundy}.} \bibinfo{year}{2020}\natexlab{}.
\newblock \showarticletitle{Checking Smart Contracts with Structural Code
  Embedding}.
\newblock \bibinfo{journal}{\emph{IEEE Transactions on Software Engineering}}
  (\bibinfo{year}{2020}).
\newblock


\bibitem[\protect\citeauthoryear{Greene and Johnstone}{Greene and
  Johnstone}{2018}]%
        {greene2018investigation}
\bibfield{author}{\bibinfo{person}{Richard Greene} {and}
  \bibinfo{person}{Michael~N Johnstone}.} \bibinfo{year}{2018}\natexlab{}.
\newblock \showarticletitle{An investigation into a denial of service attack on
  an ethereum network}.
\newblock  (\bibinfo{year}{2018}).
\newblock


\bibitem[\protect\citeauthoryear{Huang, Shihab, Xia, Lo, and Li}{Huang
  et~al\mbox{.}}{2018}]%
        {huang2018identifying}
\bibfield{author}{\bibinfo{person}{Qiao Huang}, \bibinfo{person}{Emad Shihab},
  \bibinfo{person}{Xin Xia}, \bibinfo{person}{David Lo}, {and}
  \bibinfo{person}{Shanping Li}.} \bibinfo{year}{2018}\natexlab{}.
\newblock \showarticletitle{Identifying self-admitted technical debt in open
  source projects using text mining}.
\newblock \bibinfo{journal}{\emph{Empirical Software Engineering}}
  \bibinfo{volume}{23}, \bibinfo{number}{1} (\bibinfo{year}{2018}),
  \bibinfo{pages}{418--451}.
\newblock


\bibitem[\protect\citeauthoryear{Jiang, Liu, and Chan}{Jiang
  et~al\mbox{.}}{2018}]%
        {Contractfuzzer}
\bibfield{author}{\bibinfo{person}{Bo Jiang}, \bibinfo{person}{Ye Liu}, {and}
  \bibinfo{person}{WK Chan}.} \bibinfo{year}{2018}\natexlab{}.
\newblock \showarticletitle{Contractfuzzer: Fuzzing smart contracts for
  vulnerability detection}. In \bibinfo{booktitle}{\emph{Proceedings of the
  33rd ACM/IEEE International Conference on Automated Software Engineering}}.
  ACM, \bibinfo{pages}{259--269}.
\newblock


\bibitem[\protect\citeauthoryear{Jiang, Misherghi, Su, and Glondu}{Jiang
  et~al\mbox{.}}{2007}]%
        {jiang2007deckard}
\bibfield{author}{\bibinfo{person}{Lingxiao Jiang}, \bibinfo{person}{Ghassan
  Misherghi}, \bibinfo{person}{Zhendong Su}, {and} \bibinfo{person}{Stephane
  Glondu}.} \bibinfo{year}{2007}\natexlab{}.
\newblock \showarticletitle{Deckard: Scalable and accurate tree-based detection
  of code clones}. In \bibinfo{booktitle}{\emph{29th International Conference
  on Software Engineering (ICSE'07)}}. IEEE, \bibinfo{pages}{96--105}.
\newblock


\bibitem[\protect\citeauthoryear{Kalra, Goel, Dhawan, and Sharma}{Kalra
  et~al\mbox{.}}{2018}]%
        {Zeus}
\bibfield{author}{\bibinfo{person}{Sukrit Kalra}, \bibinfo{person}{Seep Goel},
  \bibinfo{person}{Mohan Dhawan}, {and} \bibinfo{person}{Subodh Sharma}.}
  \bibinfo{year}{2018}\natexlab{}.
\newblock \showarticletitle{ZEUS: Analyzing Safety of Smart Contracts}. In
  \bibinfo{booktitle}{\emph{25th Annual Network and Distributed System Security
  Symposium (NDSS’18)}}.
\newblock


\bibitem[\protect\citeauthoryear{Kitchenham and Pfleeger}{Kitchenham and
  Pfleeger}{2008}]%
        {kitchenham2008personal}
\bibfield{author}{\bibinfo{person}{Barbara~A Kitchenham} {and}
  \bibinfo{person}{Shari~L Pfleeger}.} \bibinfo{year}{2008}\natexlab{}.
\newblock \showarticletitle{Personal opinion surveys}.
\newblock In \bibinfo{booktitle}{\emph{Guide to advanced empirical software
  engineering}}. \bibinfo{publisher}{Springer}, \bibinfo{pages}{63--92}.
\newblock


\bibitem[\protect\citeauthoryear{Li and Long}{Li and Long}{2018}]%
        {li2018detecting}
\bibfield{author}{\bibinfo{person}{Ao Li} {and} \bibinfo{person}{Fan Long}.}
  \bibinfo{year}{2018}\natexlab{}.
\newblock \showarticletitle{Detecting standard violation errors in smart
  contracts}.
\newblock \bibinfo{journal}{\emph{arXiv preprint arXiv:1812.07702}}
  (\bibinfo{year}{2018}).
\newblock


\bibitem[\protect\citeauthoryear{Li, Jiang, Chen, Luo, and Wen}{Li
  et~al\mbox{.}}{2017}]%
        {xiaoqiSurvey}
\bibfield{author}{\bibinfo{person}{Xiaoqi Li}, \bibinfo{person}{Peng Jiang},
  \bibinfo{person}{Ting Chen}, \bibinfo{person}{Xiapu Luo}, {and}
  \bibinfo{person}{Qiaoyan Wen}.} \bibinfo{year}{2017}\natexlab{}.
\newblock \showarticletitle{A survey on the security of blockchain systems}.
\newblock \bibinfo{journal}{\emph{Future Generation Computer Systems}}
  (\bibinfo{year}{2017}).
\newblock


\bibitem[\protect\citeauthoryear{Luu, Chu, Olickel, Saxena, and Hobor}{Luu
  et~al\mbox{.}}{2016}]%
        {oyente}
\bibfield{author}{\bibinfo{person}{Loi Luu}, \bibinfo{person}{Duc-Hiep Chu},
  \bibinfo{person}{Hrishi Olickel}, \bibinfo{person}{Prateek Saxena}, {and}
  \bibinfo{person}{Aquinas Hobor}.} \bibinfo{year}{2016}\natexlab{}.
\newblock \showarticletitle{Making smart contracts smarter}. In
  \bibinfo{booktitle}{\emph{Proceedings of the 2016 ACM SIGSAC Conference on
  Computer and Communications Security}}. ACM, \bibinfo{pages}{254--269}.
\newblock


\bibitem[\protect\citeauthoryear{Mossberg, Manzano, Hennenfent, Groce, Grieco,
  Feist, Brunson, and Dinaburg}{Mossberg et~al\mbox{.}}{2019}]%
        {mossberg2019manticore}
\bibfield{author}{\bibinfo{person}{Mark Mossberg}, \bibinfo{person}{Felipe
  Manzano}, \bibinfo{person}{Eric Hennenfent}, \bibinfo{person}{Alex Groce},
  \bibinfo{person}{Gustavo Grieco}, \bibinfo{person}{Josselin Feist},
  \bibinfo{person}{Trent Brunson}, {and} \bibinfo{person}{Artem Dinaburg}.}
  \bibinfo{year}{2019}\natexlab{}.
\newblock \showarticletitle{Manticore: A user-friendly symbolic execution
  framework for binaries and smart contracts}. In
  \bibinfo{booktitle}{\emph{2019 34th IEEE/ACM International Conference on
  Automated Software Engineering (ASE)}}. IEEE, \bibinfo{pages}{1186--1189}.
\newblock


\bibitem[\protect\citeauthoryear{Nakamoto}{Nakamoto}{2008}]%
        {bitcoin}
\bibfield{author}{\bibinfo{person}{Satoshi Nakamoto}.}
  \bibinfo{year}{2008}\natexlab{}.
\newblock \showarticletitle{Bitcoin: A peer-to-peer electronic cash system}.
\newblock  (\bibinfo{year}{2008}).
\newblock


\bibitem[\protect\citeauthoryear{Niemetz, Preiner, and Biere}{Niemetz
  et~al\mbox{.}}{2014}]%
        {niemetz2014boolector}
\bibfield{author}{\bibinfo{person}{Aina Niemetz}, \bibinfo{person}{Mathias
  Preiner}, {and} \bibinfo{person}{Armin Biere}.}
  \bibinfo{year}{2014}\natexlab{}.
\newblock \showarticletitle{Boolector 2.0}.
\newblock \bibinfo{journal}{\emph{Journal on Satisfiability, Boolean Modeling
  and Computation}} \bibinfo{volume}{9}, \bibinfo{number}{1}
  (\bibinfo{year}{2014}), \bibinfo{pages}{53--58}.
\newblock


\bibitem[\protect\citeauthoryear{Nikoli{\'c}, Kolluri, Sergey, Saxena, and
  Hobor}{Nikoli{\'c} et~al\mbox{.}}{2018}]%
        {Maian}
\bibfield{author}{\bibinfo{person}{Ivica Nikoli{\'c}}, \bibinfo{person}{Aashish
  Kolluri}, \bibinfo{person}{Ilya Sergey}, \bibinfo{person}{Prateek Saxena},
  {and} \bibinfo{person}{Aquinas Hobor}.} \bibinfo{year}{2018}\natexlab{}.
\newblock \showarticletitle{Finding the greedy, prodigal, and suicidal
  contracts at scale}. In \bibinfo{booktitle}{\emph{Proceedings of the 34th
  Annual Computer Security Applications Conference}}. ACM,
  \bibinfo{pages}{653--663}.
\newblock


\bibitem[\protect\citeauthoryear{Openzeppelin}{Openzeppelin}{2021a}]%
        {Pausable-Code}
\bibfield{author}{\bibinfo{person}{Openzeppelin}.} \bibinfo{year}{June.,
  2021}\natexlab{a}.
\newblock \bibinfo{booktitle}{\emph{Pausable Documentation}}.
\newblock
\urldef\tempurl%
\url{https://docs.openzeppelin.com/contracts/2.x/api/lifecycle}
\showURL{%
\tempurl}


\bibitem[\protect\citeauthoryear{Openzeppelin}{Openzeppelin}{2021b}]%
        {Pausable-Doc}
\bibfield{author}{\bibinfo{person}{Openzeppelin}.} \bibinfo{year}{June.,
  2021}\natexlab{b}.
\newblock \bibinfo{booktitle}{\emph{Pausable Documentation}}.
\newblock
\urldef\tempurl%
\url{https://github.com/OpenZeppelin/openzeppelin-contracts/blob/b0cf6fbb7a70f31527f36579ad644e1cf12fdf4e/contracts/security/Pausable.sol}
\showURL{%
\tempurl}


\bibitem[\protect\citeauthoryear{Porter}{Porter}{1980}]%
        {porterStemmer}
\bibfield{author}{\bibinfo{person}{Martin Porter}.}
  \bibinfo{year}{1980}\natexlab{}.
\newblock \bibinfo{title}{An algorithm for suffix stripping. Program, 3 (14):
  130--137}.
\newblock
\newblock


\bibitem[\protect\citeauthoryear{Sebastiani}{Sebastiani}{2002}]%
        {sebastiani2002machine}
\bibfield{author}{\bibinfo{person}{Fabrizio Sebastiani}.}
  \bibinfo{year}{2002}\natexlab{}.
\newblock \showarticletitle{Machine learning in automated text categorization}.
\newblock \bibinfo{journal}{\emph{ACM computing surveys (CSUR)}}
  \bibinfo{volume}{34}, \bibinfo{number}{1} (\bibinfo{year}{2002}),
  \bibinfo{pages}{1--47}.
\newblock


\bibitem[\protect\citeauthoryear{sklearn}{sklearn}{2021}]%
        {sklearn}
\bibfield{author}{\bibinfo{person}{sklearn}.} \bibinfo{year}{June.,
  2021}\natexlab{}.
\newblock \bibinfo{booktitle}{\emph{{scikit-learn: Machine Learning in
  Python}}}.
\newblock
\urldef\tempurl%
\url{https://scikit-learn.org/stable/}
\showURL{%
\tempurl}


\bibitem[\protect\citeauthoryear{Spencer}{Spencer}{2009}]%
        {spencer2009card}
\bibfield{author}{\bibinfo{person}{Donna Spencer}.}
  \bibinfo{year}{2009}\natexlab{}.
\newblock \bibinfo{booktitle}{\emph{Card sorting: Designing usable
  categories}}.
\newblock \bibinfo{publisher}{Rosenfeld Media}.
\newblock


\bibitem[\protect\citeauthoryear{Tsankov, Dan, Drachsler-Cohen, Gervais,
  Buenzli, and Vechev}{Tsankov et~al\mbox{.}}{2018}]%
        {Securify}
\bibfield{author}{\bibinfo{person}{Petar Tsankov}, \bibinfo{person}{Andrei
  Dan}, \bibinfo{person}{Dana Drachsler-Cohen}, \bibinfo{person}{Arthur
  Gervais}, \bibinfo{person}{Florian Buenzli}, {and} \bibinfo{person}{Martin
  Vechev}.} \bibinfo{year}{2018}\natexlab{}.
\newblock \showarticletitle{Securify: Practical security analysis of smart
  contracts}. In \bibinfo{booktitle}{\emph{Proceedings of the 2018 ACM SIGSAC
  Conference on Computer and Communications Security}}. ACM,
  \bibinfo{pages}{67--82}.
\newblock


\bibitem[\protect\citeauthoryear{Tyagi}{Tyagi}{1989}]%
        {tyagi1989effects}
\bibfield{author}{\bibinfo{person}{Pradeep~K Tyagi}.}
  \bibinfo{year}{1989}\natexlab{}.
\newblock \showarticletitle{The effects of appeals, anonymity, and feedback on
  mail survey response patterns from salespeople}.
\newblock \bibinfo{journal}{\emph{Journal of the Academy of Marketing Science}}
  \bibinfo{volume}{17}, \bibinfo{number}{3} (\bibinfo{year}{1989}),
  \bibinfo{pages}{235--241}.
\newblock


\bibitem[\protect\citeauthoryear{Vitalik and Martin}{Vitalik and
  Martin}{2021a}]%
        {EIP-3298}
\bibfield{author}{\bibinfo{person}{Buterin Vitalik} {and}
  \bibinfo{person}{Swende Martin}.} \bibinfo{year}{June., 2021}\natexlab{a}.
\newblock \bibinfo{booktitle}{\emph{EIP-3298: Removal of refunds}}.
\newblock
\urldef\tempurl%
\url{https://eips.ethereum.org/EIPS/eip-3298}
\showURL{%
\tempurl}


\bibitem[\protect\citeauthoryear{Vitalik and Martin}{Vitalik and
  Martin}{2021b}]%
        {EIP-3529}
\bibfield{author}{\bibinfo{person}{Buterin Vitalik} {and}
  \bibinfo{person}{Swende Martin}.} \bibinfo{year}{June., 2021}\natexlab{b}.
\newblock \bibinfo{booktitle}{\emph{EIP-3529: Reduction in refunds}}.
\newblock
\urldef\tempurl%
\url{https://eips.ethereum.org/EIPS/eip-3529}
\showURL{%
\tempurl}


\bibitem[\protect\citeauthoryear{Wang, Li, Lin, Ma, and Liu}{Wang
  et~al\mbox{.}}{2019}]%
        {wang2019vultron}
\bibfield{author}{\bibinfo{person}{Haijun Wang}, \bibinfo{person}{Yi Li},
  \bibinfo{person}{Shang-Wei Lin}, \bibinfo{person}{Lei Ma}, {and}
  \bibinfo{person}{Yang Liu}.} \bibinfo{year}{2019}\natexlab{}.
\newblock \showarticletitle{Vultron: catching vulnerable smart contracts once
  and for all}. In \bibinfo{booktitle}{\emph{2019 IEEE/ACM 41st International
  Conference on Software Engineering: New Ideas and Emerging Results
  (ICSE-NIER)}}. IEEE, \bibinfo{pages}{1--4}.
\newblock


\bibitem[\protect\citeauthoryear{Weili~Chen and Zhou}{Weili~Chen and
  Zhou}{2018}]%
        {ponzi-www}
\bibfield{author}{\bibinfo{person}{Jiahui Cui Edith Ngai Peilin~Zheng
  Weili~Chen, Zibin~Zheng} {and} \bibinfo{person}{Yuren Zhou}.}
  \bibinfo{year}{2018}\natexlab{}.
\newblock \showarticletitle{Detecting Ponzi Schemes on Ethereum: Towards
  Healthier Blockchain Technology}. In \bibinfo{booktitle}{\emph{Proceedings of
  the 2018 World Wide Web Conference on World Wide Web}}. International World
  Wide Web Conferences Steering Committee, \bibinfo{pages}{1409--1418}.
\newblock


\bibitem[\protect\citeauthoryear{Xia, Wan, Kochhar, and Lo}{Xia
  et~al\mbox{.}}{2019}]%
        {xia2019practitioners}
\bibfield{author}{\bibinfo{person}{Xin Xia}, \bibinfo{person}{Zhiyuan Wan},
  \bibinfo{person}{Pavneet~Singh Kochhar}, {and} \bibinfo{person}{David Lo}.}
  \bibinfo{year}{2019}\natexlab{}.
\newblock \showarticletitle{How practitioners perceive coding proficiency}. In
  \bibinfo{booktitle}{\emph{2019 IEEE/ACM 41st International Conference on
  Software Engineering (ICSE)}}. IEEE, \bibinfo{pages}{924--935}.
\newblock


\bibitem[\protect\citeauthoryear{Xiong, Zhao, Zhao, Xun, Li, Zhang, Wu, and
  Shen}{Xiong et~al\mbox{.}}{2015}]%
        {xiong2015different}
\bibfield{author}{\bibinfo{person}{Wu Xiong}, \bibinfo{person}{Qingyun Zhao},
  \bibinfo{person}{Jun Zhao}, \bibinfo{person}{Weibing Xun},
  \bibinfo{person}{Rong Li}, \bibinfo{person}{Ruifu Zhang},
  \bibinfo{person}{Huasong Wu}, {and} \bibinfo{person}{Qirong Shen}.}
  \bibinfo{year}{2015}\natexlab{}.
\newblock \showarticletitle{Different continuous cropping spans significantly
  affect microbial community membership and structure in a vanilla-grown soil
  as revealed by deep pyrosequencing}.
\newblock \bibinfo{journal}{\emph{Microbial ecology}} \bibinfo{volume}{70},
  \bibinfo{number}{1} (\bibinfo{year}{2015}), \bibinfo{pages}{209--218}.
\newblock


\bibitem[\protect\citeauthoryear{Yang and Pedersen}{Yang and Pedersen}{1997}]%
        {yang1997comparative}
\bibfield{author}{\bibinfo{person}{Yiming Yang} {and} \bibinfo{person}{Jan~O
  Pedersen}.} \bibinfo{year}{1997}\natexlab{}.
\newblock \showarticletitle{A comparative study on feature selection in text
  categorization}. In \bibinfo{booktitle}{\emph{Icml}},
  Vol.~\bibinfo{volume}{97}. Nashville, TN, USA, \bibinfo{pages}{35}.
\newblock


\bibitem[\protect\citeauthoryear{Zou, Lo, Kochhar, Le, Xia, Feng, Chen, and
  Xu}{Zou et~al\mbox{.}}{2019}]%
        {zou2019smart}
\bibfield{author}{\bibinfo{person}{Weiqin Zou}, \bibinfo{person}{David Lo},
  \bibinfo{person}{Pavneet~Singh Kochhar}, \bibinfo{person}{Xuan-Bach~D Le},
  \bibinfo{person}{Xin Xia}, \bibinfo{person}{Yang Feng},
  \bibinfo{person}{Zhenyu Chen}, {and} \bibinfo{person}{Baowen Xu}.}
  \bibinfo{year}{2019}\natexlab{}.
\newblock \showarticletitle{Smart contract development: Challenges and
  opportunities}.
\newblock \bibinfo{journal}{\emph{IEEE Transactions on Software Engineering}}
  (\bibinfo{year}{2019}).
\newblock


\end{thebibliography}

\end{document}